\documentclass[12pt,preprint]{emulateapj}
  
\usepackage{apjfonts}

\shorttitle{Rest--Frame Optical Emission in LAEs}
\shortauthors{Finkelstein et al.}

\newcommand{\sol}{$_{\odot}$}
\newcommand{\lya}{Ly$\alpha$}

\newcommand{\ergscm}{erg s$^{-1}$ cm$^{-2}$}

\newcommand{\OIII}{[O\,{\sc iii}]}
\newcommand{\NII}{[N\,{\sc ii}]}

\newcommand{\fig}[1]{Figure~\ref{#1}}
\def\arcs{\hbox{$^{\prime\prime}$}}

\def\deg{\hbox{$^\mathrm{o}$}}

\begin{document}
\slugcomment{Accepted to the Astrophysical Journal}
\title{The HETDEX Pilot Survey \sc{III}: The Low Metallicities of High-Redshift\\ Lyman Alpha Galaxies\altaffilmark{1}}

\author{Steven   L.    Finkelstein\altaffilmark{2,*}, Gary J.\ Hill\altaffilmark{3,5}, Karl Gebhardt\altaffilmark{4,5}, Joshua Adams\altaffilmark{4}, Guillermo A. Blanc\altaffilmark{4}, Casey Papovich\altaffilmark{2}, Robin Ciardullo\altaffilmark{6}, Niv Drory\altaffilmark{7}, Eric Gawiser\altaffilmark{8}, Caryl Gronwall\altaffilmark{6}, Donald P. Schneider\altaffilmark{6} \& Kim-Vy Tran\altaffilmark{2}}
\altaffiltext{1}{The data presented herein were obtained at the W.M. Keck Observatory from telescope time allocated to the National Aeronautics and Space Administration through the agency's scientific partnership with the California Institute of Technology and the University of California. The Observatory was made possible by the generous financial support of the W.M. Keck Foundation.}
\altaffiltext{2}{George P. and Cynthia Woods Mitchell Institute for Fundamental Physics and Astronomy, Department of Physics and Astronomy, Texas A\&M University, College Station, TX 77843}
\altaffiltext{3}{McDonald Observatory, Austin, TX 78712} 
\altaffiltext{4}{Department of Astronomy, University of Texas, Austin, TX 78712} 
\altaffiltext{5}{Texas Cosmology Center, University of Texas, Austin, TX 78712} 
\altaffiltext{6}{Department of Astronomy \& Astrophysics, Penn State University, State College, PA 16802}
\altaffiltext{7}{Max-Planck Institut f$\ddot\mathrm{u}$r Extraterrestrische Physik, Giessenbachstra\ss e, 85748 Garching, Germany}
\altaffiltext{8}{Department of Physics and Astronomy, Rutgers, The State University of New Jersey, Piscataway, NJ 08854}
\altaffiltext{*}{stevenf@physics.tamu.edu}

\begin{abstract}
We present the results of Keck/NIRSPEC spectroscopic observations of three \lya\ emitting galaxies (LAEs) at $z \sim$ 2.3 discovered with the HETDEX pilot survey.  We detect H$\alpha$, \OIII, and H$\beta$ emission from two galaxies at $z =$ 2.29 and 2.49, designated HPS194 and HPS256, respectively, representing the first detection of multiple rest-frame optical emission lines in galaxies at high-redshift selected on the basis of their \lya\ emission.  We find that the redshifts of the \lya\ emission from these galaxies are offset redward of the systemic redshifts (derived from the H$\alpha$ and \OIII\ emission) by $\Delta$v = 162 $\pm$ 37 (photometric) $\pm$ 42 (systematic) km s$^{-1}$ for HPS194, and $\Delta$v = 36 $\pm$ 35 $\pm$ 18 km s$^{-1}$ for HPS256.  An interpretation for HPS194 is that a large-scale outflow may be occurring in its interstellar medium.  This outflow is likely powered by star-formation activity, as examining emission line ratios implies that neither LAE hosts an active galactic nucleus.  Using the upper limits on the \NII\ emission we place meaningful constraints on the gas-phase metallicities in these two LAEs of $Z < $ 0.17 and $<$ 0.28 $Z$\sol\ (1$\sigma$).  Measuring the stellar masses of these objects via spectral energy distribution fitting ($\sim$ 10$^{10}$ and 6 $\times$ 10$^{8}$ M\sol, respectively), we study the nature of LAEs in a mass-metallicity plane.  At least one of these two LAEs appears to be more metal poor than continuum-selected star-forming galaxies at the same redshift and stellar mass, implying that objects exhibiting \lya\ emission may be systematically less chemically enriched than the general galaxy population.  We use the spectral energy distributions of these two galaxies to show that neglecting the contribution of the measured emission line fluxes when fitting stellar population models to the observed photometry can result in overestimates of the population age by orders of magnitude, and the stellar mass by a factor of $\sim$ 2.  This effect is particularly important at $z \gtrsim$ 7, where similarly strong emission lines may masquerade in the photometry as a 4000 \AA\ break.
\end{abstract}


\section{Introduction}

Star-forming galaxies at high-redshift are one of the most useful probes of the distant universe.  By studying their physical properties, we learn about the stellar mass, dust and chemical evolution during the first few Gyr after the Big Bang.  The most efficient method of learning these properties is by comparing the observed colors of galaxies (their spectral energy distributions; SEDs) to model stellar populations, finding which combination of metallicity, star formation history, extinction, age and stellar mass best matches the observed galaxies \citep[e.g.,][]{papovich01, shapley01, giavalisco02, bruzual03}.  This technique requires only broadband photometry, thus many galaxies can be studied with a single dataset.

SED fitting has been used to study thousands of galaxies at high redshift, the majority of which are either color selected \citep[Lyman break galaxies; LBGs;][]{steidel93}, or selected on the basis of a bright \lya\ emission line \citep[\lya\ emitters; LAEs; e.g.,][]{cowie98, rhoads00}.  Over 3 $< z <$ 6, LBGs have been found to be highly star-forming galaxies (star-formation rate $\sim$ 10's of M\sol\ yr$^{-1}$), with significant dust attenuation (A$_\mathrm{V} \sim$ 0.2 - 1.0 mag), stellar population ages of 100's of Myr and stellar masses of $\sim$ 10$^{10}$ -- 10$^{11}$ $M$\sol\ \citep[e.g.,][]{sawicki98, papovich01, shapley01, shapley05, yan05, yan06, eyles05, eyles07, fontana06, reddy06, huang07, overzier09, stark09}.

Over the same redshift range, LAEs appear to be less evolved in the same characteristics.  Their star-formation rates are typically more modest ($<$ 10 $M$\sol), they typically have ages $<$ 100 Myr, and similarly masses of $\lesssim$ 10$^{9}$ $M$\sol\ \citep[e.g.,][]{gawiser06b, gawiser07, pirzkal07, finkelstein07, finkelstein08, finkelstein09a, gronwall07, lai07, lai08, pentericci09, ono10a, yuma10}.  Recent results imply that at $z \geq$ 7 most galaxies have masses and ages similar to LAEs at $z <$ 6, and more evolved, LBG-like galaxies are rare \citep[e.g.,][]{ouchi09, finkelstein10a, gonzalez10, ono10b}.  While the physical properties of LBGs evolve from $z =$ 6 -- 3, the characteristics of LAEs appear unchanging, implying that they may represent newly forming galaxies at each redshift.

One of the more intriguing results to arise from high-redshift LAE stellar population analyses is the result that many LAEs appear to have dust extinction \citep[e.g.,][see also Blanc et al.\ 2010]{pirzkal07, finkelstein08, finkelstein09a, pentericci09, lai07}.  This is surprising, as dust can efficiently attenuate \lya\ photons, as they resonantly scatter with neutral hydrogen, and thus can have long path lengths prior to escaping a galaxy.  A number of scenarios have been discussed in the literature explaining how \lya\ can escape through a dusty medium, including shifting out of resonance due to outflows \citep[e.g.,][]{ahn03, verhamme08, dijkstra08, zheng10}, as well as scattering off the surfaces of dusty H\,{\sc i} clouds in a clumpy ISM \citep{neufeld91, hansen06, finkelstein08,finkelstein09a}.  These scenarios, or something similar, are likely in play, as observations of \lya\ in local star-forming galaxies have shown that \lya\ can be observed in emission even though dust is present \citep[e.g.,][]{atek08}, and both low- and high-redshift observations have found evidence that the \lya\ emission is extended \citep{nilsson09, ostlin09, finkelstein10c}.  

Nonetheless, the level of uncertainty on the dust content in high-redshift LAEs is large, as the presence of dust has been deduced by fitting the SEDs of LAEs to stellar population synthesis models.  In addition, these models place no constraints on the metallicities, as they suffer a number of degeneracies as a red spectral slope can be explained by dust, an older stellar population, or a higher metallicity.  In order to learn about these crucial evolutionary properties in more detail, rest-frame optical spectroscopy is needed.  However, at such high redshifts, these lines are shifted into the near-infrared, and thus are observationally challenging.

\citet{erb06} performed a large near-infrared spectroscopic survey of star-forming galaxies at $z \sim$ 2.3, selected via the BX method to be similar intrinsically to LBGs at $z \sim$ 3 \citep{steidel04}, using the Keck telescope with the near-infrared spectrograph NIRSPEC \citep{mclean98}.  They primarily observed in the K-band, which allowed observations of H$\alpha$ and \NII\ emission at this redshift (though they obtained H-band spectroscopy, and thus H$\beta$ and \OIII\ measurements for a few objects).  From the ratio of these lines, they derived the gas-phase metallicity of these objects.  Combining this information with stellar mass estimates, they examined the mass-metallicity relationship at $z \sim$ 2.3, finding that star-forming galaxies at that redshift followed a similar sequence of higher metallicity with higher mass as at very low redshift \citep{tremonti04}, though the sequence was shifted down by $\sim$ 0.3 dex in metallicity, showing that high-redshift star-forming galaxies are less chemically evolved than local star-forming galaxies.  A number of other studies have taken advantage of magnification due to gravitational lensing to investigate high-redshift LBG-like galaxies in detail \citep[e.g.,][]{teplitz00, finkelstein09d}.  However, until now, the physical properties of galaxies selected on the basis of \lya\ emission have not been spectroscopically probed in detail.

Observations of rest-frame optical emission lines can also  allow a determination of the systemic redshift of a galaxy.  Comparing this redshift to that of the \lya\ emission can allow one to diagnose the kinematics of the interstellar medium.  Using a composite of $\sim$ 800 LBG UV spectra, \citet{shapley03} found that \lya\ was redshifted by 360 km s$^{-1}$ with respect to the systemic redshift, and the ISM absorption lines were blueshifted by 150 $\pm$ 30 km s$^{-1}$.  \citet{steidel10} studied the ISM kinematics of a sample of 89 galaxies at $z \sim$ 2.3, finding that \lya\ and the ISM absorption lines were shifted by $+$445 and $-$164 km s$^{-1}$, respectively, when compared to the systemic redshift as measured by the H$\alpha$ emission.  These velocity differences are thought to be due to the presence of global outflows in the ISMs of these galaxies.  In such a scenario, the non-resonant nebular lines are observed at the systemic redshift, as they originate from the H\,{\sc ii} regions within the galaxy.  The ISM absorption lines are blueshifted with respect to the systemic redshift, as the absorption features are created when the outflowing gas on the near side (with respect to the observer) absorbs stellar light.  Lastly, Ly$\alpha$ preferentially escapes after it has back-scattered off of the far side of the expanding ISM, shifting the Ly$\alpha$-line center out of resonance with H\,{\sc i}, enabling it to traverse back across the galaxy, and through the near side of the ISM.  
 
Until recently, LAE samples were not available at 2 $< z <$ 2.7, which allows \OIII\ and H$\alpha$ observations with H- and K-band spectroscopy, respectively.  LAEs are typically much fainter than LBGs, thus their emission line fluxes are fainter as well, resulting in more time-demanding observations.  Recently, \citet{mclinden10} published the first rest-frame optical emission line detections from LAEs, detecting \OIII\ emission from two LAEs at $z \sim$ 3.1.  They were able to probe the kinematics of the ISM by comparing the redshifts of these lines to that of \lya, finding that both galaxies exhibited large scale outflows of 343 $\pm$ 18 and 126 $\pm$ 17 km s$^{-1}$.  Additionally, \citet{hayes10} probed the \lya\ escape fraction in high-redshift star-forming galaxies by measuring the \lya\ and H$\alpha$ fluxes with optical and near-infrared narrowband filters, though only a few objects were simultaneously detected in both lines.  However, probing further into the physical characteristics of such galaxies requires more than a single rest-frame optical line.  We present the first detections of multiple rest-frame optical emission lines from galaxies at high-redshift selected on the basis of their \lya\ emission, which we use to study in detail their physical properties.  In \S 2, we discuss the selection of our LAE sample, and our near-infrared spectroscopic observations with Keck/NIRSPEC.  In \S 3, we present our emission line measurements.  In \S 4 we measure the systemic redshift and search for the existence of outflows in the ISM, in \S 5 we probe for AGN signatures, and in \S 6 we place constraints on the dust extinction.  In \S 7, we constrain the gas-phase metallicities of our LAEs, including studying them on a mass-metallicity plane.  Lastly, in \S 8 we examine how the measured emission line fluxes can affect the SED fitting results.  We assume H$_\mathrm{0}$ = 70 km s$^{-1}$ Mpc$^{-1}$, $\Omega_\mathrm{m}$ = 0.3 and $\Omega_\mathrm{m}$ = 0.7, and unless otherwise specified we use the AB magnitude system throughout \citep{oke83}.

\section{Data}

\subsection{Sample Selection}

The Hobby Eberly Telescope Dark Energy Experiment (HETDEX) is a blind integral field spectrograph (IFS) search for LAEs at $z =$ 1.9 -- 3.5, with a primary science goal of probing dark energy via the power spectrum of 0.8 million LAEs.  The survey will begin in 2012, using the Visible Integral-field Replicable Unit Spectrograph (VIRUS) instrument, which is composed of 150 individual IFSs \citep{hill08}.  Currently, a single prototype IFS (VIRUS-P; \citet{hill08b}) is mounted on the 2.7m Harlan J. Smith Telescope at the McDonald Observatory.  Over four years, VIRUS-P has been used for 111 nights as part of a HETDEX pilot survey, discovering 103 LAEs at $z \sim$ 2--3 \citep[][hereafter Paper I and Paper II, respectively]{adams10, blanc10}.  The pilot survey probes a large volume relative to narrowband imaging searches, thus the discovered LAEs are bright, and at redshifts where well-known rest-frame optical emission lines (e.g., H$\beta$, \OIII, H$\alpha$, \NII) shift into the spectroscopic coverage of near-infrared spectrographs.

We obtained one night from NASA on the Keck II 10m telescope with NIRSPEC to obtain near-infrared spectroscopy of HETDEX pilot survey LAEs.  We selected LAEs from the pilot survey sample by requiring that the redshift be such that H$\alpha$ was observable with NIRSPEC (i.e., $z \lesssim$ 2.7) and that the \lya~line flux was greater than 10$^{-16}$ \ergscm.  The latter criterion ensured that H$\alpha$ and H$\beta$ should be bright enough to observe in 90 minutes with NIRSPEC (assuming \lya/H$\alpha$ $\sim$ 4).  Additionally, to keep the target in the NIRSPEC slit throughout an exposure, a star with K$_{Vega}$ $<$ 18 mag is required within $\sim$ 38\arcs\ of the target, such that both can be placed in the slit simultaneously.  We selected $\sim$ 10 LAEs in this manner, and chose the three brightest for observation on this run -- two in the COSMOS field \citep{scoville07}, and one in the Hubble Deep Field North \citep[HDFN;][]{dickinson03}, named HPS (HETDEX Pilot Survey) 194, HPS256 and HPS419, respectively, where the numbers correspond to the index in the HETDEX pilot survey catalog (Paper I).  These three LAEs have Ly$\alpha$-based redshifts of 2.2889, 2.4914 and 2.2357.  More details on these LAEs are listed in Table 1.

\begin{deluxetable*}{cccccccccc}
\tablecaption{Summary of NIRSPEC-Targeted LAEs}
\tablewidth{0pt}
\tablehead{
\colhead{Object} & \colhead{RA} & \colhead{Dec} & \colhead{F$_{Ly\alpha}$} & \colhead{EW$_{Ly\alpha}$} & \colhead{$\lambda_{obs}$ (Ly$\alpha$)} & \colhead{$\lambda_{corr}$ (Ly$\alpha$)} & \colhead{z$_{Ly\alpha}$} & \colhead{t$_{NIRSPEC}$ (H)} & \colhead{t$_{NIRSPEC}$ (K)}\\
\colhead{$ $} & \colhead{(J2000)} & \colhead{(J2000)} & \colhead{(10$^{-17}$ erg s$^{-1}$ cm$^{-2}$)} & \colhead{(\AA)}  & \colhead{(\AA)} & \colhead{(\AA)}& \colhead{$ $} & \colhead{(s)} & \colhead{(s)}\\
}
\startdata
HPS194&10:00:14.18&02:14:26.11&61.0 $^{+4.9}_{-4.3}$&106 $^{+15}_{-12}$&3998.0&3998.2&2.2889 $\pm$ 0.0004&5400&5400\\
HPS256&10:00:28.33&02:17:58.44&31.4 $^{+9.3}_{-6.5}$&188 $^{+66}_{-44}$&4244.1&4244.4&2.4914 $\pm$ 0.0004&1200&5400\\
HPS419&12:36:50.13&62:14:01.07&24.4 $^{+3.3}_{-5.1}$&\phantom{1}72 $^{+19}_{-18}$&3933.6&3933.5&2.2357 $\pm$ 0.0004&---&5400\\
%HPS194&150.05908&\phantom{6}2.240587
%HPS256&150.11805&\phantom{6}2.299566
%HPS419&189.20890&62.233631
\enddata
\tablecomments{The equatorial coordinates provided correspond to
  the  optical counterpart  to  the Ly$\alpha$  emission  which  were
  used as  the NIRSPEC  targets.   The detections  of  rest-frame
  optical emission lines  at the  expected  wavelengths confirm  these
  counterparts as correct (with the possible exception of HPS419).  The corrected wavelengths are shifted to the local standard of rest, to correct for both the Earth's and Sun's motions at the time of the observation.  The
  error  on the  redshift reflects  a typical  uncertainty on  the
  central  wavelength  of the Ly$\alpha$  line  of 0.5  \AA, which
  dominates over the  typical  wavelength solution error of 0.05 \AA.  The final two columns denote the NIRSPEC exposure times in each of the two filters used.}
\end{deluxetable*}

\subsection{NIRSPEC Observations}
Our observations took place on UT 21 February 2010, and conditions were photometric throughout the night.  We observed in low--resolution spectroscopy mode, with a slit width of 0.76\arcs\ and the standard slit length of 42\arcs.  Throughout the night, the cross disperser was kept at 35.76\deg.  We observed with the NIRSPEC--7 filter (hereafter referred to as the K--band), resulting in a wavelength coverage of 2.02 -- 2.43 $\mu$m, and the NIRSPEC--5 filter (hereafter referred to as the H--band), resulting in a wavelength coverage of 1.48 -- 1.76 $\mu$m.  The resolution in the K-band was R $\sim$ 1500.

We observed all three LAEs for 90 minutes in the K--band filter, obtaining 6$\times$15-minute exposures.  We also observed HPS194 for 90 minutes (6$\times$15-minutes) in the H--band, while we only obtained 20 minutes (2$\times$10 minutes) in the H--band on HPS256 due to time constraints.  Telluric standards were observed before and after each observing setup, where we selected stars from the Hipparcos survey catalog with a spectral type near F0V, which minimizes hydrogen absorption lines common in early--type stars, and metallic lines common in later--type stars.  Both science and calibration spectroscopic observations were taken in an ABBA pattern.  In addition, we obtained arc lamp calibration images between each observing setup, as the wavelength solution may drift throughout the night.  NIRSPEC is known to have significant persistence when the integrated counts pass $\sim$ 10$^{4}$ per pixel.  During the afternoon, we examined the decay of this persistence by taking dark frames, and we found it to disappear on timescales of $\sim$ 15 minutes.  However, as a precaution, we varied the position on the slit of both our objects and our standard stars, such that adjacent exposures should not have science data falling on the same detector rows.

Our targets were acquired using the ``invisible acquisition'' mode of NIRSPEC.  In this mode, an alignment star must be placed in the slit simultaneously with the object, and the star must have K$_{Vega}$ $\lesssim$ 18 to be useful.  We found alignment stars of K$_{Vega}$ = 16.5, 17.0 and 18.0 at distances of 26\arcs, 17\arcs\ and 31\arcs\ from HPS194, HPS256 and HPS419, respectively.  We first acquired the star at the center of the slit, then we slid the star toward one end of the slit, such that the star and the target LAE were equidistant from the slit center.  During an exposure, we manually guided the slit using the slit--viewing camera (SCAM), which continuously images the same field of view as the spectrograph, excepting the light which is transmitted down the slit.  Throughout the night our seeing was steady at 0.6 -- 0.9\arcs, comparable to the size of the slit, thus when the star was well centered, only the wings of the PSF were visible to either side of the slit.  If the star started to drift out of the slit, we moved the slit to keep pace, using 0.5 pixel increments ($\sim$ 0.09\arcs), ensuring that our object stayed as well-centered in the slit as possible throughout each exposure.

\subsection{Data Reduction}

We used a combination of the Keck IDL--based REDSPEC\footnote[1]{http://www2.keck.hawaii.edu/inst/nirspec/redspec.html} package along with our own custom IDL scripts to reduce the data.  Using REDSPEC for the initial steps, we processed each object/filter combination separately.  For each combination, we used calibrations taken the closest in time, consisting of four calibration star spectra and two arc lamp images (one neon and one argon).  As the flat field is not expected to vary during the night, we used flat fields taken during the afternoon for calibration.  On each combination, we first ran the REDSPEC task {\tt spatmap}, which uses the standard star spectra to compute a spatial map of the image, allowing rectification to be performed.  The task {\tt specmap} was then run, which computes the wavelength solution, where the lines are marked interactively.  Finally, the task {\tt redspec} was run, which rectifies and extracts the spectra.  However, as our objects are so faint that their continuum light is not visible, the redspec extraction was not satisfactory.  Thus, we ran {\tt redspec} to obtain the wavelength--pixel solution, but we ran the REDSPEC task {\tt rectify} to separately rectify the science images without extraction.

We rejected cosmic rays from our rectified spectra (both science and calibration) using the LACOSMIC package \citep{vandokkum01}.  The sky was then removed by subtracting adjacent image pairs, with one each in the A and B dither position, where the object was dithered by 5\arcs\ between frames (with the exception of HPS419, where the dither was only 3\arcs\ in order to keep the alignment star in the slit).  We then removed residual sky lines from these sky-subtracted images by fitting and subtracting a 15th order polynomial to each image column in the spatial direction, where the positive and negative target and alignment star spectra were masked during the fitting.

The 1D spectral extraction was performed in IDL, by extracting a 2D rectangular box centered on the spectra.  The box was chosen to be 9 pixels wide, corresponding to 1.7\arcs, or $\sim$ twice the average seeing.  The same box size was used for both the target LAEs as well as the standard stars.  During this process, the extraction region was centered based on the H$\alpha$ or [O\,{\sc iii}] $\lambda$5007 emission lines, which were visible in the individual K-- or H--band exposures.  We also verified the wavelength solution during this process to find and correct any relative shifts between the exposures (a 1.26 pixel shift was found and corrected in four K-band observations of HPS194).  These extracted regions were then summed column by column to create a 1D spectrum for each observation.  We averaged the individual 1D spectra to create the final 1D spectrum.  This procedure was repeated for each science and standard star observation\footnote[2]{For both K--band observations and the H--band observation for HPS256, one standard spectrum was significantly deviant from the other three observations; however the deviant spectrum was excluded from the analysis by the use of a median average.}.  

While we were able to make 1D spectra for HPS194 and HPS256, we were not able to detect H$\alpha$ in individual frames for HPS419.  Given the positive detection of H$\alpha$ emission in the other two LAEs, we would have expected a detection here.  However, the alignment star used with HPS419 was very faint, with K$_{Vega}$ = 18, thus we required 180 s SCAM integrations to see the star in the SCAM image.  Given the level of slit shift seen in the other objects with much shorter SCAM exposures and the large distance between HPS419 and its alignment star, we believe that it is likely that HPS419 drifted out of the slit multiple times, which resulted in the non-detection.  Future observations of very faint targets are advised to use brighter alignment stars (K$_\mathrm{Vega}$ $\leq$ 17.5) to avoid this issue.  For the remainder of the paper, we focus on HPS194 and HPS256, which have significant emission line detections.

The spectra of these two LAEs were corrected for Telluric absorption as well as flux calibrated using the standard star spectra.  This was done by taking a \citet{kurucz93} model star spectrum of the same spectral type as the standard star, and normalizing its flux such that the H or K--band magnitude of the model matched that of the standard, where the photometry of the standard was obtained from the Two Micron All Sky Survey (2MASS) Point Source Catalog.  We computed a calibration array by taking the ratio of the normalized model spectrum to the observed standard spectrum, interpolating over absorption features common to both spectra.  This calibration array was multiplied into the observed object spectrum, thus performing the flux calibration and correcting for Telluric absorption.  In addition, as our targets are unresolved  from the ground, this calibration also corrects for point--source slit losses, as the standard and object had the same spectral extraction size, and the standard spectrum was normalized to match the total magnitude for that star.

A noise spectrum was created for each exposure for each object.  This was done by first extracting a series of regions of the same size as used on the object, starting at the bottom of the 2D spectrum, and moving successively up by 9 rows at a time (i.e., so that no two rows are included in more than one noise estimate) in every input image, providing a 1D spectrum for each of the extracted rows.  The noise in each spectral pixel was obtained by fitting a Gaussian to the pixel values in each column, performing two iterations of 3$\sigma$ clipping to reject any real objects, and then measuring the standard deviation.  This process created a 1D error spectrum for each image.  These arrays were then added in quadrature, and divided by the number of input frames (i.e., error propagation of a mean) to create a final 1D error spectrum for each object.  Although this spectrum provides a good approximation of the photometric error at each wavelength, it does not include uncertainties on the flux calibration, which are important as some of the physical quantities which we wish to calculate rely on ratios between the two filters we have observed.  To include this effect, during the extraction of the standard star we computed a calibration error spectrum from the standard deviation of the standard star spectra in each wavelength bin.  We found that the noise from this process was typically much less ($\lesssim$ 1\%) than the photometric noise; nonetheless this error spectrum was then added in quadrature to the photometric error spectrum for each object to include the uncertainties in the calibration process in our final error spectrum.  We note that we have not included any uncertainty due to drifting of the standard star during the calibration exposures, as these were short and thus this error is likely minimal.

\begin{figure*}
\epsscale{0.55}
\vspace{4mm}
\plotone{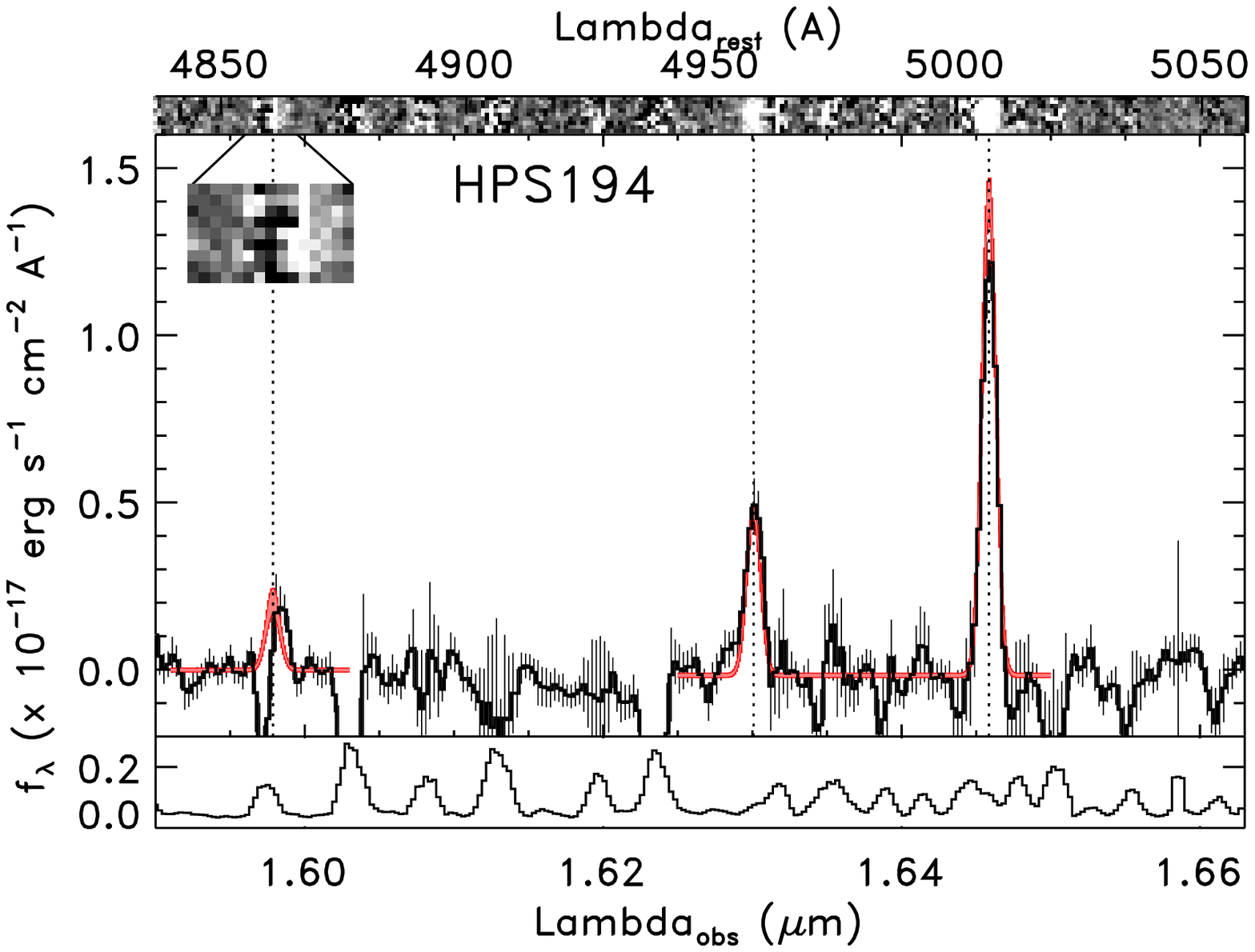}
\plotone{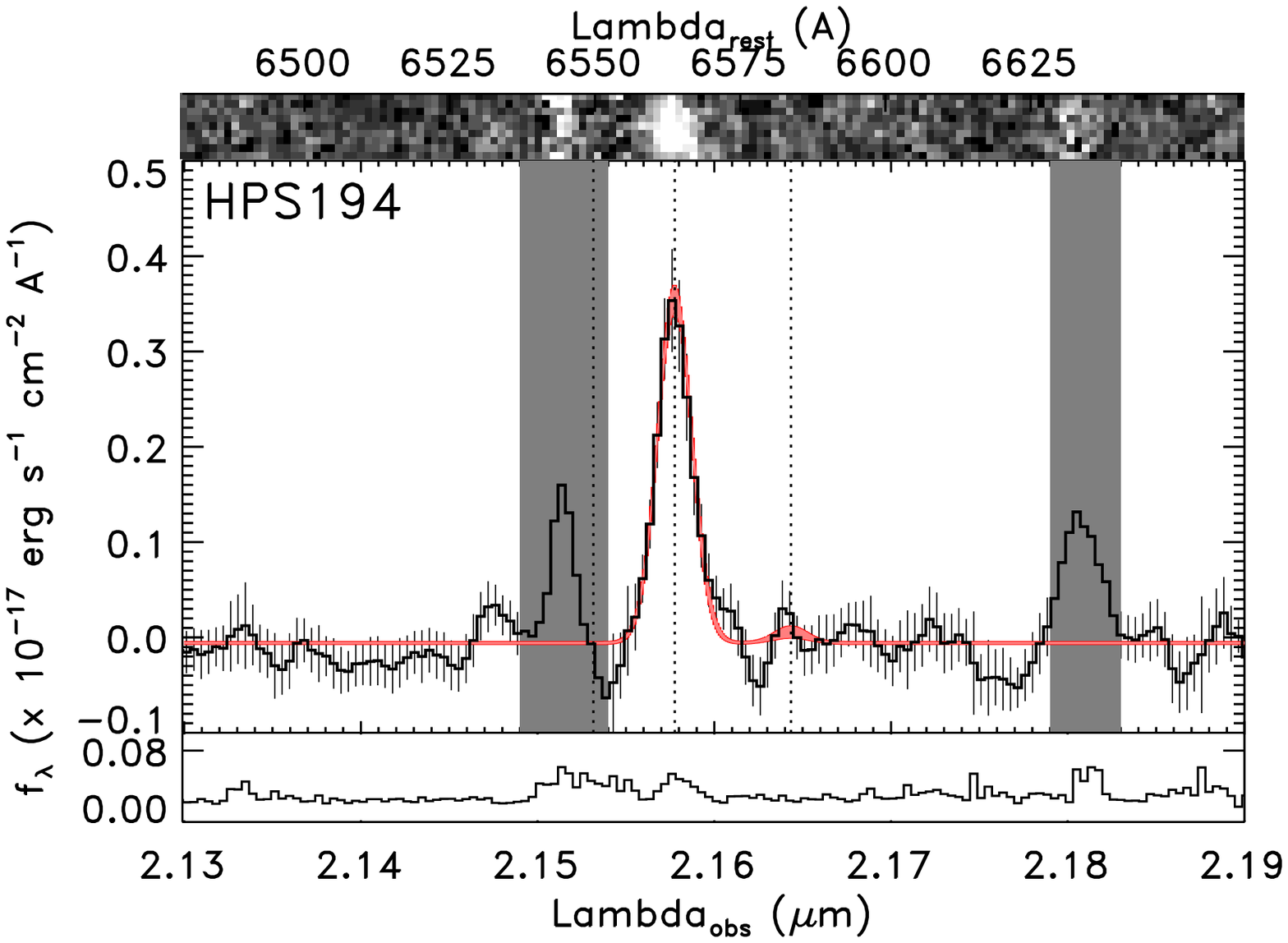}
\vspace{4mm}
\caption{Left: H-band spectrum for HPS194.  The top panel shows the reduced two-dimensional spectrum.  In the middle panel, the black line shows the observed 1D spectrum, where for display purposes we have boxcar smoothed it to match the NIRSPEC resolution at these wavelengths.  Based on the redshift of \lya\ for this object, we easily identify both \OIII\ emission lines.  We also identify H$\beta$ emission, although the blue half of the line is affected by a sky line residual.  The red line denotes the best fit Gaussians to these emission lines, where the width of the line denotes the 1$\sigma$ range of allowable fits.  We fixed the redshift and FWHM of H$\beta$ to match \OIII\ $\lambda$5007 to allow us to fit a Gaussian to half of a line.  The bottom panel shows the 1D error spectrum, in the same units as the middle panel, magnified on the flux axis to show the structure of the noise.  Right:  K-band spectrum for HPS194.  While H$\alpha$ is obviously strongly detected, \NII\ only appears as a weak bump, and is not significant.  Strong sky-line residuals are masked out by gray vertical bars.}\label{cosmos16}
\end{figure*}

\begin{figure*}
\epsscale{0.55}
\vspace{4mm}
\plotone{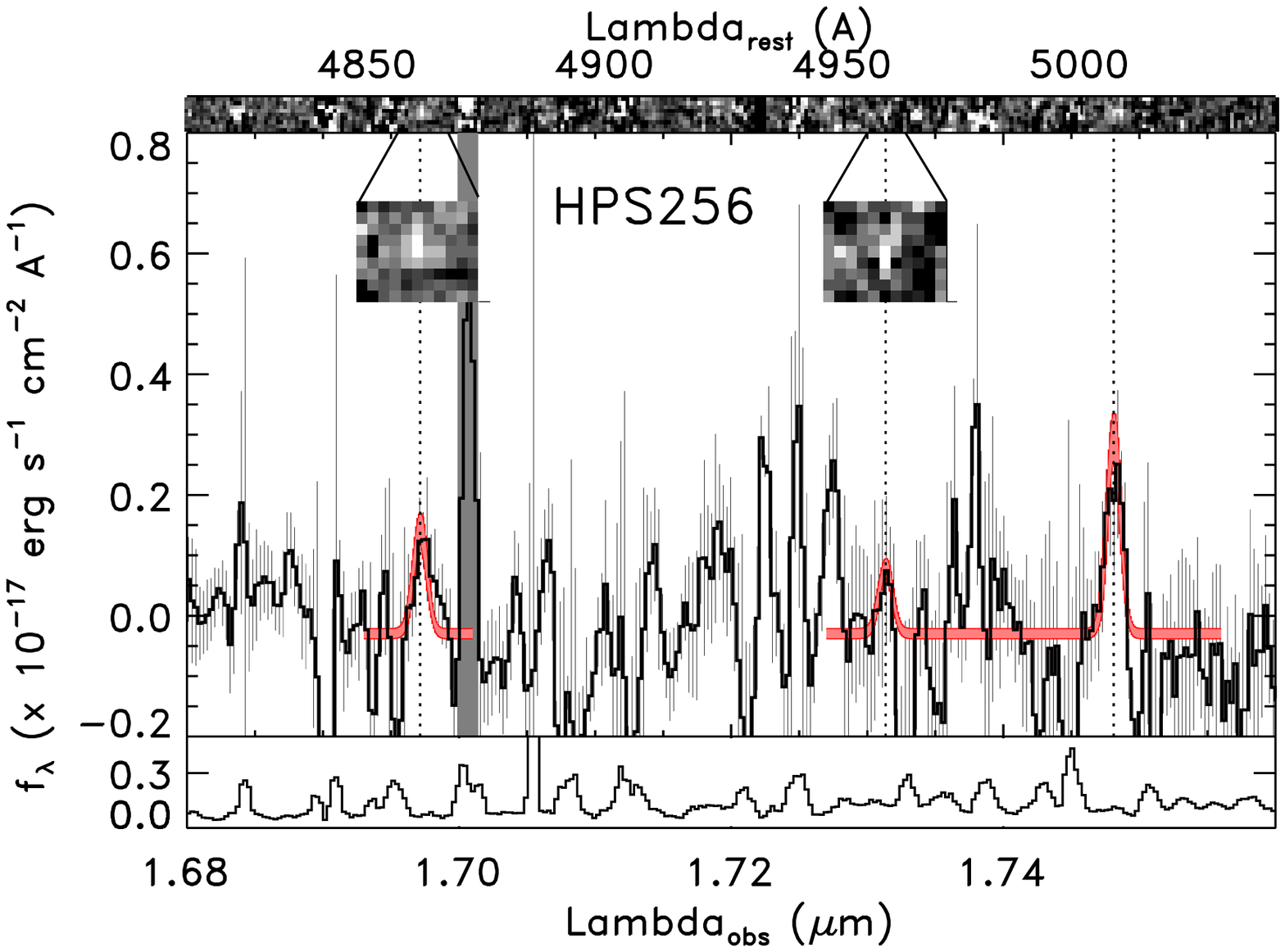}
\plotone{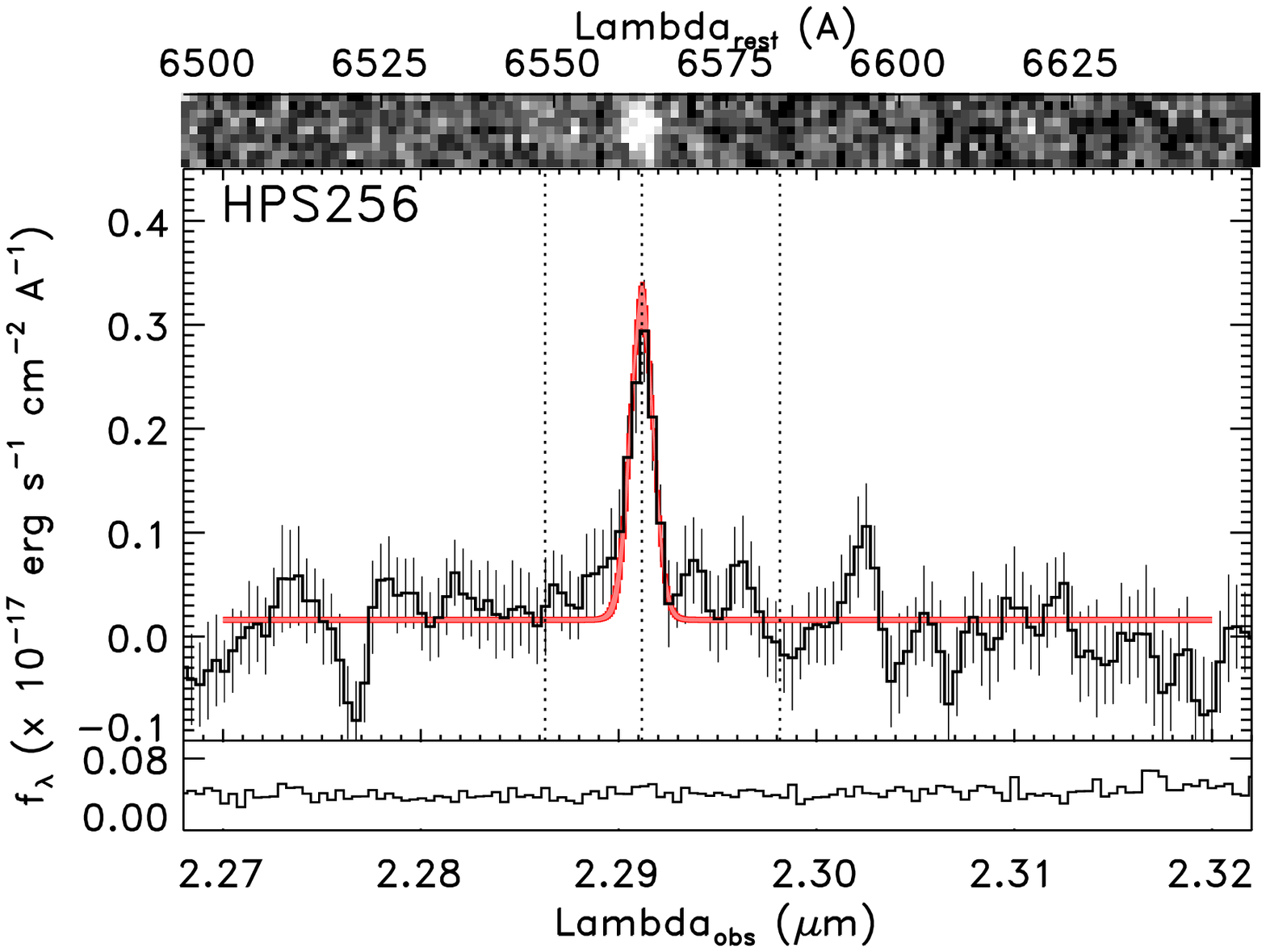}
\vspace{4mm}
\caption{Left: H-band spectra for HPS256, with the lines and shaded regions the same as in \fig{cosmos16}.  The redshift and FWHM of the \OIII\ $\lambda$4959 and H$\beta$ lines were fixed to the value measured for \OIII\ $\lambda$5007.  This observational setup only received 20 minutes of exposure time, thus this spectrum is relatively noisy.  Right:  K-band spectrum for HPS256.  Although we do not see \NII\ emission by eye, we still allow a line (restricted to have a non-negative line flux) to be fit to its expected position to obtain an estimate of the upper limit on its line flux.}\label{cosmos10}
\end{figure*}

\section{Emission Line Measurements}

\fig{cosmos16} and \fig{cosmos10} show the reduced and flux-calibrated 1D spectra for HPS194 and HPS256, respectively, with the 1D error spectra shown in the bottom of each panel.  The left and right panels in each figure show the region of the spectra around H$\beta$ and \OIII\ (left) and H$\alpha$ and \NII\ $\lambda$6583 (right), using the measured redshift of Ly$\alpha$ to estimate the expected positions of these rest-frame optical lines.  In \fig{cosmos16}, it is apparent we have detected H$\alpha$ emission in HPS194 at high significance.  We associate the weak positive feature at $\lambda$ = 2.164 $\mu$m as \NII.  In the left panel, both \OIII\ $\lambda$4959 and \OIII\ $\lambda$5007 are strongly detected.  There also appears to be an emission line at the expected position of H$\beta$.  Although the blue portion of this line coincides with a night sky line, the right portion is unaffected.  One can see the red half of the H$\beta$ emission line when examining the 2D spectra, thus we conclude it is real.  In \fig{cosmos10}, we again see a strong detection of H$\alpha$ in HPS256, though there is no positive feature at the expected position of \NII.  In the H-band spectrum, we see possible emission lines at the expected positions of \OIII\ 4959, \OIII\ 5007 and H$\beta$, though it is difficult to tell if these lines are significant, given the increased noise present due to the shorter exposure time of this observation.

In order to measure the properties of the emission lines, we fit a Gaussian curve to each detected line using the MPFIT IDL software package\footnote[3]{http://www.physics.wisc.edu/$\sim$craigm/idl/fitting.html}.  For the K-band spectra for both objects, we fit a double Gaussian around the position of the detected H$\alpha$ flux, allowing it to fit one Gaussian to the H$\alpha$ flux, and a second Gaussian to \NII.  We fit a 500 -- 600 \AA-wide region of the spectra to allow ample room for the estimation of the continuum.  MPFIT requires as input an estimate of the fitted parameters, which are the continuum flux, line wavelength, line full-width at half-maximum (FWHM) and line flux.  These numbers were estimated from the 1D spectra.

\begin{deluxetable*}{cccccccc}
\tablecaption{Measured Emission Line Properties}
\tablewidth{0pt}
\tablehead{
\colhead{Object} & \colhead{Line} & \colhead{$\lambda_{rest}$} & \colhead{$\lambda_{obs}$} & \colhead{$\lambda_{corr}$} & \colhead{z} & \colhead{F$_{line}$} & \colhead{EW$_{rest}$}\\
\colhead{$ $} & \colhead{$ $} & \colhead{(\AA)} & \colhead{(\AA)} & \colhead{(\AA)} & \colhead{$ $} & \colhead{(10$^{-17}$ erg s$^{-1}$ cm$^{-2}$)} & \colhead{(\AA)}\\
}
\startdata
HPS194&H$\beta$&4861&15978.7 $\pm$ 0.1&15978.2&---&\phantom{1}2.43 $\pm$ 0.43&\phantom{2}33 $\pm$ 6\\
HPS194&[O\,{\sc iii}]&4959&16300.8 $\pm$ 0.1&16300.3&---&\phantom{1}5.67 $\pm$ 0.48&\phantom{2}66 $\pm$ 6\\
HPS194&[O\,{\sc iii}]&5007&16458.6 $\pm$ 0.2&16458.1&2.28702 $\pm$ 0.00003&16.90 $\pm$ 0.48&202 $\pm$ 6\\
HPS194&H$\alpha$&6563&21577.6 $\pm$ 0.4&21577.0&2.28767 $\pm$ 0.00005&\phantom{1}8.80 $\pm$ 0.27&177 $\pm$ 5\\
\smallskip
HPS194&[N\,{\sc ii}]&6583&21643.4 $\pm$ 0.4&21642.8&---&$<$ 0.15&$<$ 3\\
HPS256&H$\beta$&4861&16971.3 $\pm$ 0.8&16970.8&---&\phantom{1}2.24 $\pm$ 0.31&---\\
HPS256&[O\,{\sc iii}]&4959&17313.5 $\pm$ 0.8&17313.0&---&$<$ 0.47&---\\
HPS256&[O\,{\sc iii}]&5007&17481.1 $\pm$ 0.8&17480.6&2.49123 $\pm$ 0.00017&\phantom{1}4.11 $\pm$ 0.47&---\\
HPS256&H$\alpha$&6563&22911.7 $\pm$ 0.44&22911.0&2.49094 $\pm$ 0.00007&\phantom{1}4.44 $\pm$ 0.29&---\\
HPS256&[N\,{\sc ii}]&6583&22981.5 $\pm$ 0.44&22980.8&---&$<$ 0.18&---\\
\enddata
\tablecomments{The redshifts of H$\beta$ and [O\,{\sc iii}] $\lambda$4959 were fixed to that of [O\,{\sc iii}] $\lambda$5007, and the redshifts of [N\,{\sc ii}] were fixed to that of H$\alpha$.  Both [N\,{\sc ii}] lines as well as \OIII\ $\lambda$4959 in HPS256 were detected at $<$ 3 $\sigma$ significance, thus we list the 1 $\sigma$ upper limits.  Using the redshifts listed above, the weighted mean redshifts for HPS194 and HPS256 are $z =$ 2.28712 $\pm$ 0.00002 and $z =$ 2.49098 $\pm$ 0.00006, respectively.  The equivalent widths were computed from the ratio of the observed line flux to that of the continuum flux near the line from the best-fit model.  HPS256 is not detected in the IRAC bands, thus the EWs for the emission lines in that object are not well constrained and so are not listed.}
\end{deluxetable*}

As the \NII\ line is either only weakly detected, or not detected at all, we forced the second Gaussian to have the same FWHM and redshift as the first Gaussian for both objects, thus the only free parameter for \NII\ is the line flux.  This was performed iteratively, by forcing the FWHM and redshift of the \NII\ fit to be fixed to initially the same value as was estimated for H$\alpha$, and then in successive iterations forcing the \NII\ parameters to be equal to the H$\alpha$ FWHM and redshift from the previous iteration.  This was done until the difference between the two lines was $\Delta z <$ 0.00001 and $\Delta$FWHM $<$ 0.01 \AA.  In addition, the flux of the \NII\ line was constrained to be positive, as this forbidden line cannot be seen in absorption.

Similar Gaussians were fit to the H-band spectra for both objects.  As these lines are more separated then H$\alpha$+\NII\, we fit separate Gaussians to H$\beta$ and the two \OIII\ lines.  We fixed the ratio of the \OIII\ F$_{5007}$/F$_{4959}$ to equal the theoretical ratio of 2.98 \citep{storey00}.  We fixed both the redshift and the FWHM of the weaker H$\beta$ and \OIII\ $\lambda$4959 lines to match that of the \OIII\ $\lambda$5007 line during the fitting.  For HPS194, a portion of the H$\beta$ line was lost due to sky emission, thus the much greater noise on the blue half of H$\beta$ ensures that the red half (where we see the significant flux) dominates the fit.  For all fits, the emission line flux was constrained to be non-negative.

It is difficult to judge the detection significance of a given emission line simply by examining the spectra around the line, as surrounding regions might be heavily affected by residual sky lines, while the line region itself might be relatively free.  We thus quantified the uncertainties in the emission line fit parameters by running a series of 10$^{3}$ Monte Carlo simulations on each spectrum.  In each simulation, every spectral element was varied by a Gaussian random deviate proportional to the flux error at that wavelength.  The altered spectrum then had its emission lines fit in the same way as was done on the actual data (i.e., holding the same parameters fixed which were with the data).  Through these simulations, we thus compiled 10$^{3}$ estimates of the line wavelengths (and thus redshifts), FWHMs and fluxes.  We computed the 1$\sigma$ uncertainty on these parameters as one half of the spread of the central 68\% of these values.  The simulation results are shown in \fig{cosmos16} and \fig{cosmos10} as the width of the red line, which shows the central 68\% of the simulation fits, highlighting the 1$\sigma$ uncertainty on these fits.  In Table 2 we tabulate all of the measured emission line properties, as well as the uncertainties on these properties from our simulations.  We find that \NII\ is not significantly detected in either object, and \OIII\ $\lambda$4959 is not detected in HPS256.  We detect the remaining lines at more than 5$\sigma$ significance.

In \S 7 we use upper limits on the \NII\ line fluxes to place constraints on the metallicities.  To confirm our original estimate of the 1$\sigma$ uncertainties on the \NII\ lines, we performed another set of simulations where we input into the K-band spectra of each object mock emission lines of varying strengths at the position of \NII.  We ran Monte Carlo simulations on each mock spectrum to assess the signal-to-noise ratio of the mock emission line detection.  Plotting the signal-to-noise versus the mock line flux, we fit a line to the data, where the slope of this line is the 1$\sigma$ flux limit.  We found 1$\sigma$ limits on the \NII\ flux of 1.47 $\times$ 10$^{-18}$ and 1.75 $\times$ 10$^{-18}$ erg s$^{-1}$ cm$^{-2}$ for HPS194 and HPS256, respectively.  We use these limits in the following analysis.

\section{Results}

\subsection{Redshift and Velocity Offsets}
The simple fact that we detect H$\alpha$ emission (along with H$\beta$ and \OIII) near the expected wavelength confirms the original identification of our two objects as LAEs at $z \approx$ 2.3 and 2.5.  However, by comparing the exact values of the redshift measured from Ly$\alpha$ to that measured from the rest-frame optical emission lines, we can examine the kinematic state of the interstellar medium (ISM) in these galaxies.  In many LBGs Ly$\alpha$ has been found to have a slightly higher redshift than the systemic redshift of the galaxy by $\sim$ 300-400 km s$^{-1}$, which coupled with measurements of blueshifted absorption lines implies that outflows may be present in the ISMs of these galaxies \citep[e.g.,][]{shapley03, steidel10}.  

Although these outflows appear ubiquitous in LBGs, the data did not exist until recently to probe for velocity offsets between \lya\ and the systemic redshifts in the typically less luminous population of LAEs.  \citet{mclinden10} recently measured \OIII\ emission in two $z \sim$ 3.1 LAEs.  They found that the Ly$\alpha$ emission lines in their galaxies were redshifted by 343 $\pm$ 18 and 126 $\pm$ 17 km s$^{-1}$ with respect to the \OIII\ emission, which presumably hails from the systemic redshift.  Although this is a small sample, if these offsets are due to outflows, then the outflow velocities in one of these galaxies is smaller than those in LBGs.  This result could easily be explainable, as LBGs form stars at a higher rate \citep[e.g.,][]{papovich01, shapley01, yan05, stark09}, which could produce more intense outflows.  However, a larger sample of LAE velocity offsets needs to be compiled before strong conclusions can be made.

With our emission line redshifts, we have the data necessary to probe velocity offsets in our LAEs.  However, the optical and NIR datasets were taken on different dates and at different observatories.  The relative motion of the Earth (both rotational and orbital) as well as the motion of the Sun with respect to the observed target will thus be slightly different for the two datasets.  Thus, before we compare the redshifts of the emission lines, we need to correct the observed line wavelengths for this effect.  We computed the necessary correction from the observed data to the local standard of rest using the online V$_{LSR}$ calculator\footnote[4]{http://fuse.pha.jhu.edu/support/tools/vlsr.html}.  The optical spectra of HPS194 and HPS256 were obtained at McDonald Observatory on UT 2008-12-29 and 2008-11-29, respectively.  Including the time of observation, we found that the radial velocity of the observer with respect to the target was $-$15.55 and $-$21.37 km s$^{-1}$, respectively.  The NIR spectra of these two objects were obtained at the Keck Observatory on UT 2010-02-21.  Similarly including the observation times, the radial velocity with respect to the target was 8.66 and 8.40 km s$^{-1}$ for the HPS194 H- and K-band spectra, and 9.09 and 8.94 km s$^{-1}$ for the HPS256 H- and K-band spectra, respectively.  The sign of these measurements is such that a positive number means that the observer was moving away from the target (with respect to V$_{LSR}$) at the time of the observation, so the line would appear at a slightly higher redshift.

We used these velocities to correct the observed emission line wavelengths in both datasets, and then we computed the redshifts of the lines, both of which are listed in Tables 1 and 2.  For the NIR data, we only computed the redshift when it was a free parameter, as in both objects the redshift of \NII\ was fixed to match that of H$\alpha$, and in HPS256, the redshift of H$\beta$ and \OIII\ $\lambda$4959 was fixed to match that of \OIII\ $\lambda$5007.  We computed the systemic redshift of each galaxy using the weighted mean of all measured rest-frame optical emission line redshifts.  

The systemic redshift for HPS194, using H$\alpha$ and both \OIII\ lines, is $z_{sys} =$ 2.28712 $\pm$ 0.00002.  For HPS256, using H$\alpha$ and \OIII\ $\lambda$5007, it is $z_{sys} =$ 2.49098 $\pm$ 0.00006.  The \lya\ based redshift for these two galaxies are $z_{Ly\alpha} =$ 2.2889 $\pm$ 0.0004 and 2.4914 $\pm$ 0.0004, respectively.  Comparing these two redshifts, we find that for HPS194, \lya\ is redshifted compared to the systemic redshift by $\Delta$v = 162 $\pm$ 37 km s$^{-1}$.  For HPS256, \lya\ is redshifted compared to $z_{sys}$ by $\Delta$v = 36 $\pm$ 35 km s$^{-1}$.  However, the error on the weighted mean redshift does not account for systematic differences between the values used to compute the mean.  For example, in HPS194, the redshifts of the H$\alpha$ and \OIII\ $\lambda$5007 lines are significantly different.  These velocity differences are well within the instrumental resolution, thus we conclude that they are a systematic associated with the wavelength calibration.  We add a systematic error term to the above result, where the systematic error is the standard deviation of the velocity offsets from the individual lines used to compute the mean.  Our final velocity offset results are thus: $\Delta$v = 162 $\pm$ 37 (photometric) $\pm$ 42 (systematic) km s$^{-1}$ for HPS194, and $\Delta$v = 36 $\pm$ 35 $\pm$ 18 km s$^{-1}$ for HPS256.

These offsets may be due to large-scale outflows in the ISMs of these galaxies, although only the result for HPS194 is statistically significant.  Compared to the outflows in LBGs, the measured velocity offsets in our LAEs appear lesser in magnitude, similar to what was found in \citet{mclinden10} for one of their two LAEs.  While a sample of four objects is insufficient to make robust conclusions, it is intriguing that most LAEs appear to have smaller outflow velocities than LBGs.  This could be due to a variety of physical effects, as we still do not truly understand the difference between these two populations of galaxies.  However, in the literature, LAEs tend to be less luminous, lower in stellar mass, forming stars at a lesser rate, and living in smaller dark matter halos than LBGs \citep[e.g.,][]{papovich01, shapley03,gawiser06b,gawiser07,pirzkal07, gronwall07, finkelstein09a,stark09,finkelstein10a, ono10a}.  \citet{steidel10} analyzed their sample of $z \sim$ 2.3 galaxies to discern if the outflow velocities they derived were correlated with any physical property, including those stated above.  They found that the velocity differences between the systemic redshift and the interstellar absorption features and \lya\ emission in their sample do not positively correlate with any of the physical properties.  This is surprising, as it has been shown that the outflow speed can correlate with the star-formation rate \citep{martin05, rupke05}.  \citet{steidel10} did find that the velocity offset of the ISM absorption features anti-correlates with both the baryonic and dynamical masses (though at less than 3$\sigma$), which could imply that the more massive galaxies are accreting inflowing material at the systemic velocity.  Thus, it appears that more work is needed to probe the root physical cause of the outflows in high-redshift star-forming galaxies.  We note that internal absorption of the blue half of the \lya\ emission line can result in the observed redshift of \lya\ being slightly higher than systemic without invoking outflows \citep[e.g.,][]{zheng10, laursen10}.  Future deep optical spectroscopy can measure the redshifts of absorption lines in the ISM in these LAEs, which can be used as a second probe for the existence of outflows.

\section{Signatures of Active Galactic Nuclei}
Although star formation can produce \lya\ emission, active galactic nuclei (AGNs) also produce copious amounts of ionizing photons, and thus their host galaxies emit in \lya\ as well.  Understanding whether the production of \lya\ photons is dominated by star formation or AGN activity is crucial in understanding not only the emission line, but also the characteristics of the galaxy as well.  At high-redshift ($z >$ 3), X-ray emission and high-ionization state emission lines (i.e., C\,{\sc iv} $\lambda$1549; He\,{\sc ii} $\lambda$1640; C\,{\sc iii}] $\lambda$1909) are used to diagnose the presence of AGNs.  Using these methods, the AGN fractions of LAEs at $z >$ 3 have been found to be typically 0 -- 2\% \citep[e.g.,][]{malhotra03, wang04, gawiser06b, ouchi08}.  The fraction of AGNs among LAEs appears to rise towards lower redshifts, with an AGN-fraction of $\sim$ 4 -- 5\% at $z \sim 2.1$ \citep{nilsson09, guaita10}, and up to 40\% at $z \sim 0.3$ \citep{finkelstein09c, scarlata09, cowie10}.  While this may indicate that lower-redshift LAEs are more likely to host AGNs, it could also indicate that less luminous, or obscured AGNs are lurking in LAEs at high redshift.  This is especially critical at the redshifts of our LAE sample here, as they lie near to the redshift of peak AGN activity \citep[e.g.,][]{osmer82, richards06, silverman08}.  

At lower redshifts, the mid-infrared fluxes of galaxies can be used to diagnose the presence of AGNs, as they typically exhibit red power law slopes \citep{stern05, donley08}.  However, this typically requires detections in multiple mid-infrared bands, which is not always possible.  Locally, AGNs have been identified using the line ratio diagram of \citet[][hereafter BPT]{baldwin81}, which separates galaxies into star-forming and AGN sequences, independent of X-ray luminosity.  This separation is done by plotting the ratio of \OIII/H$\beta$ versus \NII/H$\alpha$, where AGNs experience elevated levels of both ionized metal lines with respect to star-forming dominated galaxies due to the harder UV spectra of AGNs.  

\begin{figure}[t!]
\epsscale{1.15}
\plotone{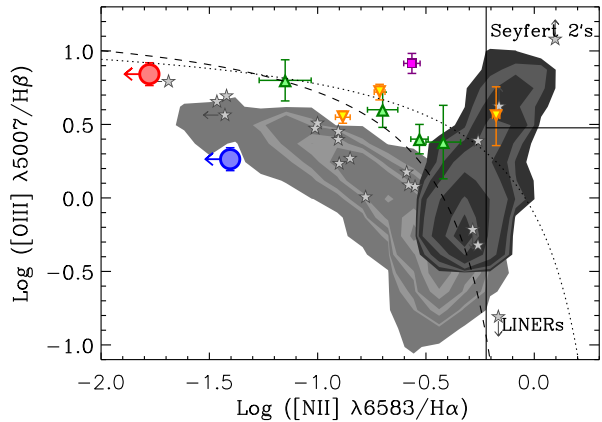}
\caption{Our two LAEs are plotted on a BPT line-ratio diagnostic diagram, with HPS194 in red, and HPS256 in blue.  The light and dark gray contoured regions denote where local star-forming and AGN-dominated galaxies lie, respectively, using data from the SDSS.  The dotted line is the maximum starburst curve from \citet{kewley01}, while the dashed curve is an updated SF/AGN demarcation line from \citet{kauffmann03c}. The solid lines denote regions expected to be inhabited by Seyfert 2's and LINERs \citep{kauffmann03c}.  Green triangles denote four $z \sim$ 2.3 LBGs studied by \citet{erb06}.  The purple square is the position in this diagram of the z=2.73 lensed galaxy the 8 o'clock arc \citep{finkelstein09d}, while the yellow inverted triangles represent three $z \sim$ 2 lensed galaxies studied by \citet{hainline09}.  The gray stars denote $z \sim$ 0.3 LAE-analogs \citep{finkelstein09e}.}\label{bpt}
\end{figure}

To study the possibility of AGN activity in our sample of LAEs, we first analyzed both the COSMOS {\it Chandra} Bright Source Catalog \citep[v2.1][]{puccetti09} and the COSMOS {\it XMM} Point-like Source Catalog for X-ray counterparts near to our LAEs.  In both catalogs (which have flux limits of $\sim$ 5 $\times$ 10$^{-16}$ erg s$^{-1}$ cm$^{-2}$ in the {\it XMM} soft-band and {\it Chandra} full-band, respectively), we found no X-ray counterparts within 10\arcs\ of our LAEs.  We then plotted the positions of our two LAEs on a BPT diagram, shown in \fig{bpt}.  We denote the limits due to the upper limit on the \NII\ fluxes by arrows.  We plot contours of where galaxies from the Sloan Digital Sky Survey \citep[SDSS;][]{york00} fall in this plane\footnote[5]{We used line fluxes from a sample of $\sim$ 10$^{5}$ SDSS star-forming galaxies and $\sim$ 3 $\times$ 10$^{3}$ SDSS AGNs from the CMU-Pitt Value-Added Catalog: http://nvogre.phyast.pitt.edu/vac/}, with the AGN-dominated objects shown as the dark-gray contours, and the star-formation dominated objects shown as the light-gray contours.  Both of our LAEs lie far from the AGN sequence, with their limits possibly pushing them even further, thus we conclude that the ionization in these objects is dominated by star-formation activity.  

Examining our two objects, HPS256 appears consistent with the local star-forming sequence, while HPS194 lies above and to the left, close to the maximum starburst curve of \citet{kewley01}.  A similar trend of enhanced \OIII/H$\beta$ ratios is seen in many star-forming galaxies at $z \sim$ 2.  \citet{erb06} studied four $z \sim$ 2 LBGs in this plane, and found that even the ones far from the AGN sequence appear enhanced in \OIII/H$\beta$ relative to the SDSS star-forming galaxies (these may also be enhanced in \NII/H$\alpha$, moving them diagonally off the star-forming sequence).  This effect was also seen in lensed LBGs at similar redshifts studied by \citet{finkelstein09d} and \citet{hainline09}.  Lastly, enhanced \OIII/H$\beta$ ratios have also been noted in the $z \sim$ 2.3 galaxy BX418 \citep{erb10} which exhibits weak high-ionization emission in their rest-frame UV, similar to the $z \sim$ 2.5 galaxy MTM 095355--545428 which also does not appear to host AGN activity \citep{malkan96}.  These observations together imply that the H\,{\sc ii} regions in high-redshift galaxies have a stronger ionization field than seen locally, which could be due to a reduced metallicity or an increased electron density\citep[e.g.,][]{liu08, erb06, brinchmann08}.  Further data are needed to determine if elevated \OIII/H$\beta$ ratios extend to the majority of LAEs, and to diagnose whether it is purely linked to star formation.  However, we conclude that there are no indications that either LAE in our sample hosts an AGN.

\section{Hydrogen Line Diagnostics}

\subsection{Dust Extinction}
We can use the spectroscopic measurements of H$\alpha$ and H$\beta$ emission to compute the Balmer decrement in our LAEs, which is a measure of how much the H$\alpha$/H$\beta$ ratio has been increased due to differential dust attenuation of the bluer line.  To perform this measurement, we assume an intrinsic ratio of H$\alpha$/H$\beta$ = 2.86, following \citet{osterbrock89} for the conditions in a typical H\,{\sc ii} region (and Case B recombination).  Assuming the starburst dust extinction law of \citet{calzetti00}, we compute the color excess due to dust extinction, finding E(B-V) = 0.20 $\pm$ 0.15 (A$_\mathrm{V}$\footnote[6]{Assuming a Calzetti et al.\ (2000) extinction curve, the conversion from E(B-V) to A$_\mathrm{V}$ is a factor of $\sim$ 4.} = 0.8 $\pm$ 0.6 mag) and E(B-V) $<$ 0.13 (A$_\mathrm{V}$ $<$ 0.5 mag) for HPS194 and HPS256, respectively.  The color excess for HPS256 is formally negative, so we quote the 1$\sigma$ upper limit here.  We note that similar negative values have been seen before \citep[e.g.,][]{atek09}, and may be indicative of H$\beta$ photons scattering off of a reflection nebula within the galaxy.  

The color excess is very sensitive to changes in the Balmer decrement, thus the $<$ 10$\sigma$ H$\beta$ detections result in a high uncertainty on the color excess, thus HPS194 is consistent with A$_\mathrm{V}$ = 0 as well as A$_\mathrm{V} \sim$ 1 mag.  We note that these measurements ignore any stellar absorption, which should be a small effect here \citep{rosagonzalez01}.  A 1 \AA\ correction to the H$\beta$ EW results in an uncertainty on E(B-V) of $\sim$ 0.05, much smaller than our current uncertainties.  We conclude that higher fidelity H$\beta$ measurements are needed to make robust measurements of the dust extinction levels in LAEs.  Paper II used the UV spectral slope to estimate the level of dust extinction in these objects, and found E(B-V) = 0.09 $\pm$ 0.05 (A$_\mathrm{V}$ = 0.4 $\pm$ 0.2 mag) for HPS194, and E(B-V) = 0.10 $\pm$ 0.10 (A$_\mathrm{V}$ = 0.4 $\pm$ 0.4 mag) for HPS256, in broad agreement with our Balmer decrement measurements. 

\subsection{\lya\ Escape Fraction}
We can still gain some insight into the ISMs in LAEs by comparing the H$\alpha$ flux to the \lya\ flux from these two objects from the HETDEX pilot survey.  Previous results show a range of \lya/H$\alpha$ of $\sim$ 1 -- 9 in LAEs at $z \sim$ 0.3 and $z \sim$ 2.2 \citep{finkelstein10b, hayes10}.  We find the ratio of \lya\ flux to H$\alpha$ flux in our objects to be less than the Case B value of 8.7, at 6.9 $\pm$ 0.5 in HPS194, and 7.1 $\pm$ 1.1 in HPS256\footnote[7]{Although \lya\ and H$\alpha$ were obtained with two different instruments, no ``slit'' correction should be necessary.  \lya\ was obtained with an IFS and thus is the total \lya\ flux, while H$\alpha$ was corrected to total via the flux calibration (i.e., the standard star spectrum was scaled to match the 2MASS magnitude; see \S 2.3).}.  This result immediately shows that dust if most likely present, as both ratios are significantly less than the Case B value, although as mentioned above, we cannot place strict limits on the exact value of the extinction in these objects.  We acknowledge that the IGM can also attenuate the \lya\ flux, though significant attenuation is unlikely at $z \sim$ 2 (average IGM opacity at $z =$ 2.3 would attenuate just 5\% of a \lya\ line).

If the dust extinction is very low in both objects, this implies that the ISM is very uniform, as only a small amount of dust is significantly reducing the flux ratio from the intrinsic Case B value.  In the case of HPS194, if the extinction is near the measured value of A$_\mathrm{V}$ = 0.8 mag, this implies an inhomogeneous ISM, as this level of extinction would attenuate \lya\ significantly in a homogeneous geometry.  The specific value of A$_\mathrm{V}$ = 0.8 mag in this object would necessitate $\tau_{Ly\alpha} < \tau_{continuum}$, implying that the \lya\ equivalent width (EW) we observe is actually enhanced over the intrinsic value \citep{finkelstein09a}.  However, the large uncertainties on the dust extinctions in these galaxies renders us unable to make conclusions about the ISM, other than that a full range of geometries appear possible.  The result that both objects exhibit \lya/H$\alpha$ $\sim$ 7 should prove useful to future surveys targeting H$\alpha$ emission from luminous LAEs.

Irrespective of the ISM geometry, we can estimate the fraction of \lya\ photons which are escaping these galaxies by comparing the observed \lya/H$\alpha$ ratio to the Case B value.  This would imply f$_{esc}$(\lya) = 0.80 $\pm$ 0.06 for HPS194 and 0.81 $\pm$ 0.13 for HPS256.  However, we know dust is present, thus these estimates are upper limits.  To compute the true escape fraction of \lya\ photons, we must correct the H$\alpha$ flux for dust attenuation.  We do this using the values of E(B-V) from Paper II, which are more robust than our own Balmer decrement measurements for the reasons stated above.  We assume that dust affects nebular emission the same as the stellar continuum, as evidence for a preferentially higher attenuation on the nebular emission \citep{calzetti00} is not typically seen at high redshift \citep[e.g.,][]{erb06, hainline09}, though see also \citet{forster09}.  We find corrected \lya\ escape fractions of f$_{esc}$(\lya) = 0.61 $\pm$ 0.11 for HPS194 and f$_{esc}$(\lya) = 0.60 $\pm$ 0.22 for HPS256.  These values are consistent with those of Paper II, who find  f$_{esc}$(\lya) = 0.6 $^{+0.4}_{-0.3}$ for HPS194 and f$_{esc}$(\lya) $>$ 0.3 for HPS256 by comparing the observed \lya\ star-formation rates to those derived from the extinction-corrected rest-frame UV continuum.  However, they are much higher than the median value of f$_{esc}$(\lya) = 0.22 from Paper II, implying that these two LAEs have among the highest \lya\ escape fractions among galaxies at $z \sim$ 2--3.  This is to be expected, as the galaxies with the brightest \lya\ fluxes were selected for NIR spectroscopic followup.

\subsection{Star Formation Rates}
We use our H$\alpha$ observations measure the star-formation rates of these two LAEs via the conversion from \citet{kennicutt98}, which assumes a Salpeter initial mass function (IMF).  Our observed H$\alpha$ fluxes correspond to L$_{H\alpha}$ = 3.5 and 2.2 $\times$ 10$^{42}$ erg s$^{-1}$, corresponding to lower limits on the observed SFRs of 28.0 $\pm$ 0.8 and 17.3 $\pm$ 0.9 M\sol\ yr$^{-1}$ for HPS194 and HPS256, respectively.  Correcting the H$\alpha$ luminosities for dust attenuation as in the previous sub-section, we find dust-corrected SFRs of 36.8 $\pm$ 5.8 for HPS194, and 23.6 $\pm$ 6.5 M\sol\ yr$^{-1}$ for HPS256.  Even without a dust correction, these SFRs are much larger than are typically seen in LAEs at $z >$ 3 \citep[e.g.,][]{gronwall07}, although similar to some bright LAEs studied at $z =$ 2.25 \citep{nilsson10}.  However, these high SFRs are again likely due to selection effects, as we selected the galaxies with the brightest \lya\ lines from our sample for NIR spectroscopic followup to increase the chances of a H$\alpha$ detection, thus we also selected those with the highest SFRs.  Additionally, as we show below, although our galaxies have \lya\ in emission, they have masses consistent with continuum-selected star-forming galaxies at these redshifts, which typically have SFRs $>$ 10 M\sol\ yr$^{-1}$ at L$^{\ast}$ \citep[e.g.,][]{reddy06}.  Future work with a more homogeneous sample of $z \sim$ 2 LAEs will place stronger constraints on whether they are forming stars at a higher rate than LAEs at higher redshift.

\section{Gas-Phase Metallicity}

Although the presence of dust implies that many LAEs are not composed of pristine gas, they may be more metal poor than LBGs or other continuum-selected star-forming galaxies (i.e., BX selected) as they are less evolved in their other physical properties.  With our data, we can place constraints on the metallicities of high-redshift \lya-selected galaxies using both the N2 and O3N2 gas-phase metallicity index calibrations of \citep{pettini04}.  However, \NII\ emission is not detected in either of our LAEs, though given the expected strength of the \NII\ emission line from the observed H$\alpha$ flux, this is not surprising as an AGN-dominated ionizing spectrum would be required to produce a detectable amount of \NII\ emission.  As such, we cannot measure direct metallicity values, but we can place meaningful constraints.

Using the 1$\sigma$ upper limit on the \NII\ flux with the N2 index we place upper limits on the gas-phase metallicities in our galaxies of $Z$ $<$ 0.17 $Z$\sol\ for HPS194, and $Z$ $<$ 0.28 $Z$\sol\ for HPS256 (2$\sigma$ upper limits are $Z$ $<$ 0.25 $Z$\sol\ and $Z$ $<$ 0.41 $Z$\sol, respectively).  The 1$\sigma$ upper limits from the O3N2 index are similar, with $Z$ $<$ 0.17 $Z$\sol\ and  $Z$ $<$ 0.34 $Z$\sol\ for these two LAEs, respectively.
 
\begin{figure*}[t!]
\epsscale{1.00}
\plotone{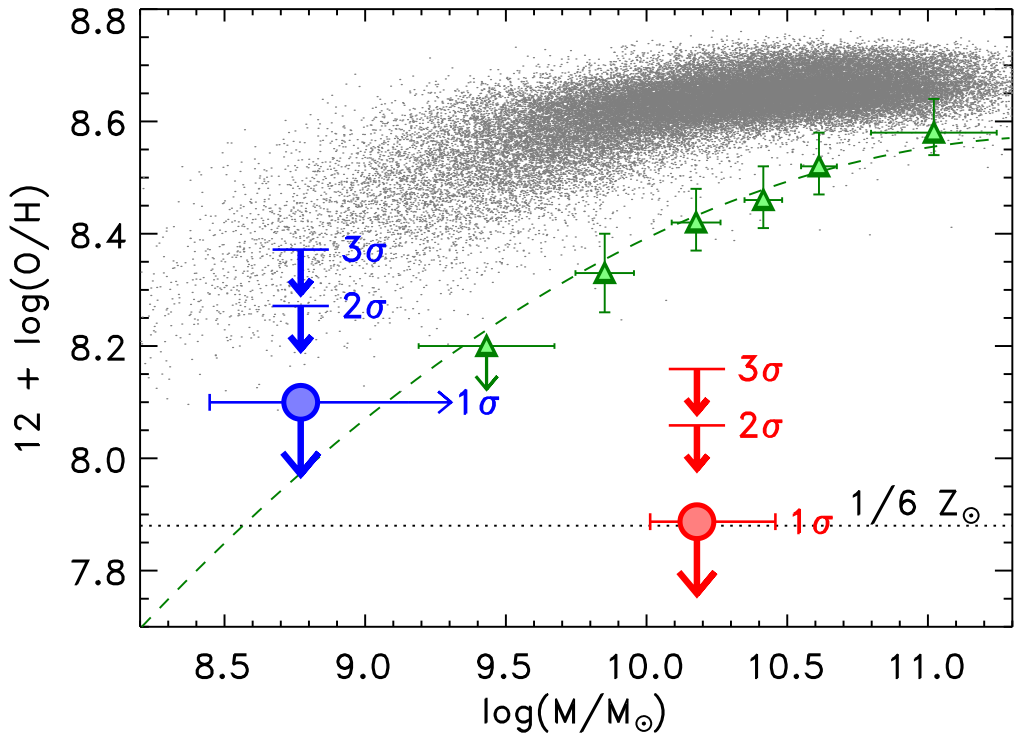}
\caption{The positions of our two LAEs on a stellar mass -- gas-phase-metallicity plane, where red represents HPS194 and blue represents HPS256.  The small gray circles denote the SDSS sample of \citet{tremonti04}, which represents low-redshift, star-forming galaxies, with the metallicity re-derived with the N2 index to match the measurement used for our objects.  The green triangles denote the $z \sim$ 2.3 star-forming galaxy sample of \citet{erb06}, which comprises 87 LBG-like (BX-selected) galaxies split into six bins of stellar mass.  The arrows for each LAE denote the 1, 2 and 3$\sigma$ confidence limits on the metallicities of our two LAEs, while the horizontal error bar on the 1$\sigma$ points denote the 68\% confidence ranges on their stellar masses.  The stellar masses are those from the emission-line corrected fits discussed in \S 8.  The amount of stellar mass in old stars in HPS256 is not well-constrained, thus this object may be up to 100$\times$ more massive (see \S 8.1).  The upper limit on the metallicity of HPS194 is much lower than the $z \sim$ 2.3 LBG $M-Z$ trend, even though it has $\sim$ L$^{\ast}_{UV}$ luminosity, implying that galaxies exhibiting \lya\ in emission may be more metal-poor than the general galaxy population at the same redshift.  The dotted line denotes $Z$ = 1/6 $Z$\sol.  This is the metallicity of BX418 from \citet{erb10}, which is the most metal-poor galaxy known at high redshift.}\label{mz}
\end{figure*}

These constraints on the metallicities of these objects are extremely interesting.  The metallicity of HPS194 is constrained to be less than all but one previously measured high-redshift galaxy.  \citet{erb10} recently published a detailed study of the $z =$ 2.3 BX galaxy Q2343-BX418 (hereafter BX418).  BX418 appears unreddened, with a stellar mass of $\sim$ 10$^{9}$ M\sol, and a H$\alpha$-derived SFR of 15 M\sol\ yr$^{-1}$.  BX418 also has \lya\ in emission, with EW$_{rest}$ = 56 \AA, a \lya\ escape fraction of $\sim$ 40\%, and a somewhat broad \lya\ profile with FWHM $\sim$ 850 km s$^{-1}$ (the \lya\ emission of HPS194 has a similar width).  They found an upper limit on its metallicity from the N2 index of $Z_{N2}$ $<$ 0.28 $Z$\sol.  They then used the knowledge of its low metallicity to use the lower branch of the R23 calibration, finding an R23 metallicity of $Z_{R23}$ = 0.17 $Z$\sol.   Thus BX418 appears broadly similar to one or both of our LAEs in many physical properties.

To place these metallicity limits in context, we compare them to the stellar masses of these galaxies, which we derive in the following section using publically available photometry in the COSMOS field.  Using the emission-line corrected SED-fitting results, we find stellar masses of 1.5 (1.0 -- 2.9) $\times$ 10$^{10}$ $M$\sol\ and 5.9 (2.8 - 14.2) $\times$ 10$^{8}$ $M$\sol\ for HPS194 and HPS256, respectively (the values in parenthesis represent the 68\% confidence ranges on the stellar masses).  \fig{mz} shows these two LAEs plotted on a mass-metallicity ($M-Z$) relation.  We compare their positions on this plane to those of low-redshift star-forming galaxies from the SDSS taken from \citet{tremonti04}, as well as LBGs at $z \sim$ 2.3 from \citet{erb06}.  Metallicities for both comparison samples have been recalculated using the N2 index.  LBGs at $z \sim$ 2.3 lie below the low-redshift mass-metallicity sequence, though they follow the same trend, as lower-mass galaxies have lower metallicities.  

Using the metallicity limits of our objects, we study our two high-redshift LAEs in this plane.  HPS256 is consistent with the $z \sim$ 2.3 LBG $M-Z$ relation at 1$\sigma$.  However, as we discuss below, the uncertainty on the stellar mass of HPS256 is high as it is undetected in the photometry at $\lambda$ $>$ 1 $\mu$m, thus it may contain up to $\sim$100$\times$ more stellar mass.  If this is the case than it may reside below the $z \sim$ 2.3 LBG $M-Z$ relation.  HPS194 has tighter constraints on its stellar mass, and an even lower limit on its metallicity.  We find that it lies below the $z \sim$ 2.3 LBG $M-Z$ relation by at least 0.3 dex (at 2$\sigma$).  

A higher ionizing field in the H\,{\sc ii} regions of high redshift galaxies is a possibility given the observed offsets in the BPT diagram of many high-redshift galaxies, including HPS194.  The N2 metallicity index we use was calibrated from local star-forming galaxies, which do not have higher-than-normal emission line ratios.  Thus, the locally derived N2 index may not be directly applicable to high-redshift galaxies.  The most direct way to show this would be to re-derive the N2 index with a high-redshift sample.  However, this will require the measurement of the \OIII\ $\lambda$4363 auroral emission line over a large sample, and the faint nature of this line makes this unfeasible with current observatories.  However, the R23 index calibration of \citet{kobulnicky99} takes the ionization parameter into account, and thus may be a better alternative.  This is more difficult observationally, as at these redshifts it will not only require J and H-band spectroscopy to detect the desired lines, it also requires a robust measurement of the Balmer decrement (requiring K-band spectroscopy), such that the emission lines can be corrected for dust extinction.  We do not have J-band spectroscopy of our LAEs, however \citet{erb10} have computed both R23 and N2 for the $z =$ 2.3 BX418 galaxy, which appears to have very similar properties to the LAEs in our sample.  As discussed above, \citet{erb10} derive a metallicity for BX418 of Z = 0.17 Z\sol\ with the R23 index, and Z $<$ 0.28 Z\sol\ with the N2 index.  The two indices produce consistent values of the metallicity, thus in this case at least, the offset in the BPT diagram does not appear to be affecting the N2-based metallicity measurements.  Finally, even if the N2 index systematically shifts with an increased ionization parameter, the offset of at least one of our LAEs from the LBG mass-metallicity relation is real, as \citet{erb06} also used the N2 index to derive the metallicities of their sample.

As discussed above, SED-fitting studies have shown that LAEs are likely less evolved than LBGs in stellar mass, stellar population age and dust extinction.  Our above results show that the metallicities in at least one of our two LAEs are less than what one may expect from their stellar masses, thus LAEs may be less evolved in metallicity than LBGs, possibly by a large amount.  This result is even more intriguing given the relatively large stellar mass of HPS194.  From the observed photometry, this galaxy's rest-frame luminosity corresponds to $\sim$ 1.4 L$^{\ast}$ using the $z \sim$ 2.3 UV luminosity function of \citet[][the luminosity of HPS256 corresponds to $\sim$ 0.5 L$^{\ast}$]{reddy09}.  Thus, HPS194 is as luminous and massive as a typical color-selected LBG, yet it has a much lower metallicity.  This implies that galaxies exhibiting \lya\ in emission may be more metal-poor than the general galaxy population at the same redshift, consistent with results found for a sample of low-redshift galaxies emitting in \lya\ \citep{finkelstein10b}.  Although more spectroscopic observations of high-redshift LAEs are needed before strong conclusions are possible, if this result holds and LAEs are systematically more metal poor then the LBG population at a constant mass, it could confirm that LAEs are the progenitors of $\sim$ L$^{\ast}$ LBGs at subsequent redshifts.  
%This may be likely, as the low-metallicity galaxies we have observed were selected to be among the brightest galaxies in our sample, thus it seems logical that less-luminous LAEs will have low metallicities as well.

\begin{deluxetable*}{cccccccccc}
\tablecaption{Broadband Photometry of the Sample}
\tablewidth{0pt}
\tablehead{
\colhead{Object} & \colhead{$B$} & \colhead{$V$} & \colhead{$r^{\prime}$} & \colhead{$i^{\prime}$} & \colhead{$z^{\prime}$} & \colhead{$J$} & \colhead{$K_{s}$} & \colhead{$3.6$} & \colhead{$4.5$}\\
\colhead{$ $} & \colhead{(mag)} & \colhead{(mag)} & \colhead{(mag)} & \colhead{(mag)} & \colhead{(mag)} & \colhead{(mag)} & \colhead{(mag)} & \colhead{(mag)} & \colhead{(mag)}\\
}
\startdata
HPS194&23.92 (0.06)&24.07 (0.06)&24.10 (0.06)&24.18 (0.07)&23.90 (0.12)&22.73 (0.18)&22.66 (0.14)&22.59 (0.04)&22.69 (0.08)\\
HPS256&24.93 (0.09)&25.27 (0.11)&25.39 (0.11)&25.52 (0.14)&25.47 (0.34)&$>$ 25.17&$>$ 25.18&$>$ 24.01&$>$ 23.32\\
\enddata
\tablecomments{The optical and near-infrared data is from the COSMOS Intermediate and Broadband Photometry Catalog, while the mid-infrared data is from the S-COSMOS catalog.  Magnitude 1$\sigma$ errors are listen in parentheses.  All magnitudes are corrected to total using the aperture corrections from their respective catalogs.  When we correct these fluxes for the measured Ly$\alpha$ line fluxes, we find corrected magnitudes for HPS194 of B = 24.22, and for HPS256 of B = 25.60.  Likewise, correcting the measured K$_{s}$-band flux for the measured H$\alpha$ emission, we find K$_{s}$ = 22.84 in HPS194.  HPS256 is not detected from the J-band redward, thus we list the 1$\sigma$ upper limit for the J- and K$_{s}$-bands, while we list the 5$\sigma$ upper limit for the IRAC bands, as only objects brighter than these limits were included in the S-COSMOS catalog.}
\end{deluxetable*}

\section{Effect of Emission Lines on SED-Fitting Results}
\begin{figure*}[t!]
\epsscale{0.55}
\plotone{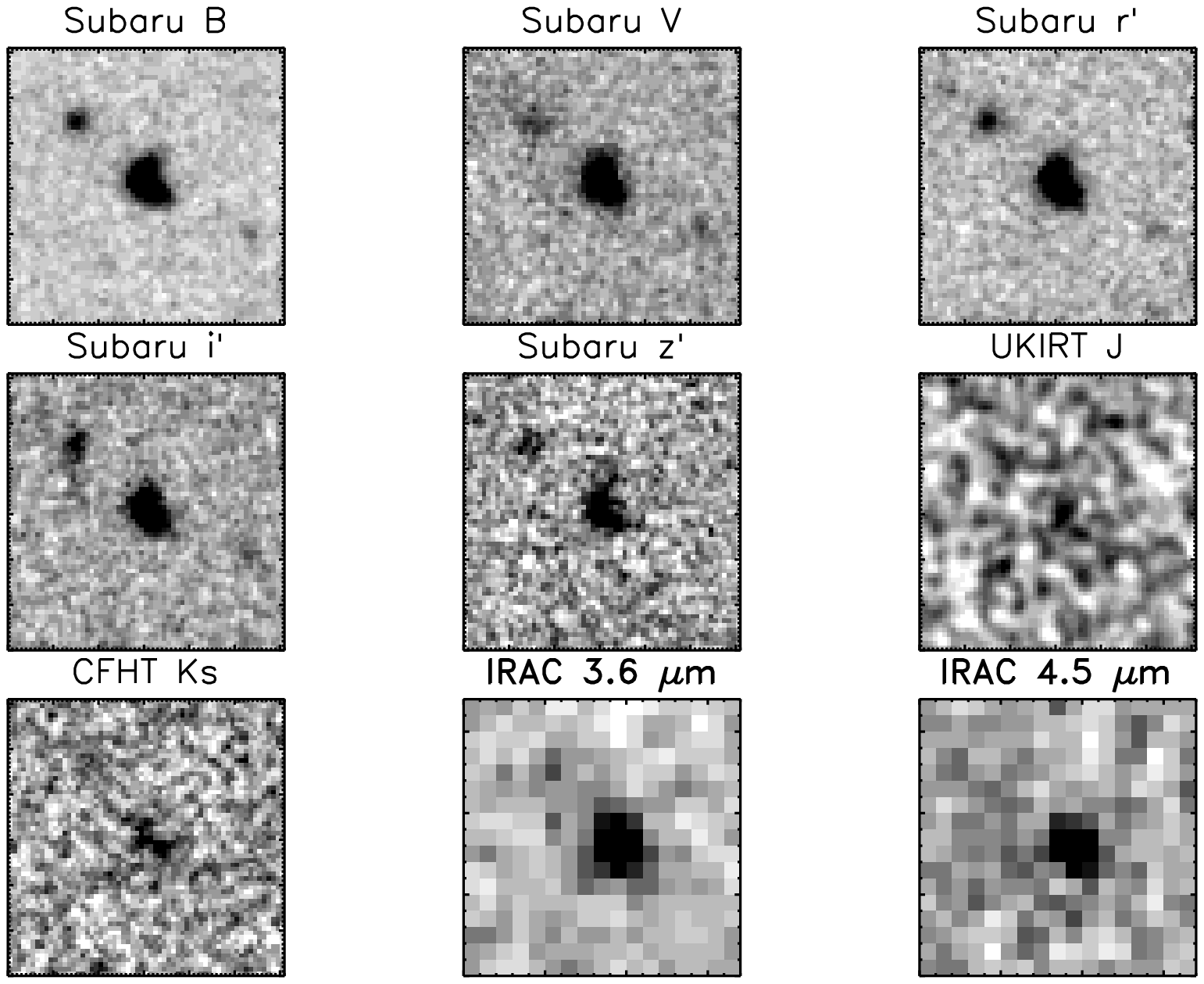}
\hspace{3mm}
\plotone{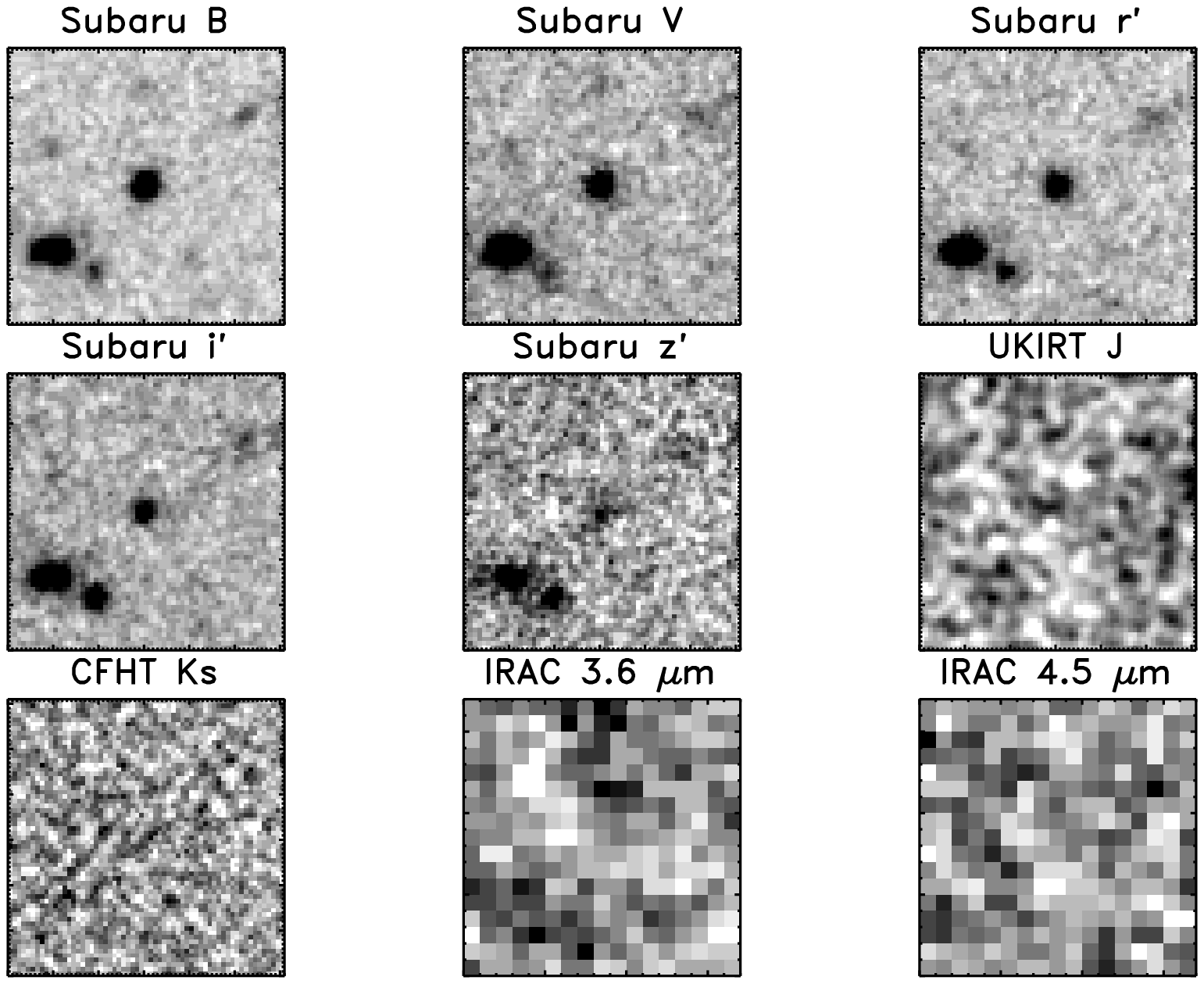}
\caption{Cutout stamps (9\arcs\ on a side) of HPS194 (left) and HPS256 (right) in the 9 bands that were used for SED fitting.  HPS256 is undetected in all bands redward of 1 $\mu$m.}\label{stamps}
\end{figure*}

\citet{mclinden10} observed that the K-band flux from their two $z \sim$ 3.1 LAEs can be wholly accounted for from their measured \OIII\ emission line fluxes, and could thus significantly affect SED-fitting results.  With our sample of LAEs, we can empirically examine how the presence of strong emission lines in the rest-frame optical affects the physical properties derived from SED fitting, using the strengths of actual measured lines.  To perform this analysis, we used the COSMOS Intermediate and Broad Band Photometry Catalog, released in April 2009\footnote[8]{http://irsa.ipac.caltech.edu/data/COSMOS/datasets.html}.  This catalog provides the photometry performed on PSF-matched images in a 3\arcs\ aperture, as well as an aperture correction to adjust to total magnitudes.  For our LAEs, we used the photometry from this catalog in the following bands (we neglected the u$^{*}$-band as it is mostly blueward of 1216 \AA\ in both objects, and thus is subject to line-of-sight IGM variations): Subaru B (0.45 $\mu$m), V (0.55 $\mu$m), r$^{\prime}$ (0.63 $\mu$m), i$^{\prime}$ (0.77 $\mu$m), z$^{\prime}$ (0.89 $\mu$m); UKIRT J (1.25 $\mu$m) and CFHT WIRCAM K$_{s}$ (2.15 $\mu$m).  Additionally, deep {\it Spitzer Space Telescope} Infrared Array Camera (IRAC) data is available from the S-COSMOS  IRAC Photometry Catalog at 3.6 and 4.5 $\mu$m \citep{sanders07}.  We used the S-COSMOS flux from a 1.9\arcs\ aperture, using the provided aperture corrections to adjust to total fluxes.  Although HPS256 was undetected in all IRAC bands, only objects with $\geq$ 5$\sigma$ detections were included in the S-COSMOS catalog \citep{sanders07}.  Thus, for these bands we only penalized the $\chi^2$ when the model flux violated the 5$\sigma$ limits.  Additionally, while the catalog has a formal 2.9$\sigma$ detection in the K$_{s}$ band for HPS256, there is no significant flux apparent in the image (see \fig{stamps}).  We thus set the flux in this band to zero for HPS256, keeping the 1$\sigma$ error as the uncertainty (the stellar mass derived when leaving this K$_{s}$ flux in is consistent within 1--2$\sigma$ of our results below).  The photometry for our sample is listed in Table 3, and 9\arcs\ cutout stamps in each band are shown in \fig{stamps}.

The photometry from these nine bands were compared to a suite of synthetic stellar populations, using the updated (2007 version) of the models of \citet[][hereafter CB07]{bruzual03}.  Models were created over a range of metallicity (0.02 -- 2.5 $Z$\sol), star-formation history (SFH; $\tau$ = 10$^{5}$ -- 10$^{9.6}$ yr), age (1 Myr -- t$_{H}[z]$) and dust extinction (A$_{V}$ = 0 -- 1.6 mag, using the extinction law of \citet{calzetti00}).  IGM attenuation was applied via the prescription of \citet{madau95}.  The best-fit model was found via $\chi^2$-minimization, with the stellar mass being calculated as a normalization between the best-fit model and the observed photometry.  We added a 5\% systematic uncertainty in quadrature to the photometric errors in each band to account for possible zeropoint uncertainties, aperture corrections, or other uncertainties which scale multiplicatively with the flux.  Uncertainties on the derived properties were obtained via 10$^3$ Monte Carlo simulations.

\begin{figure*}[t!]
\epsscale{1.05}
\plottwo{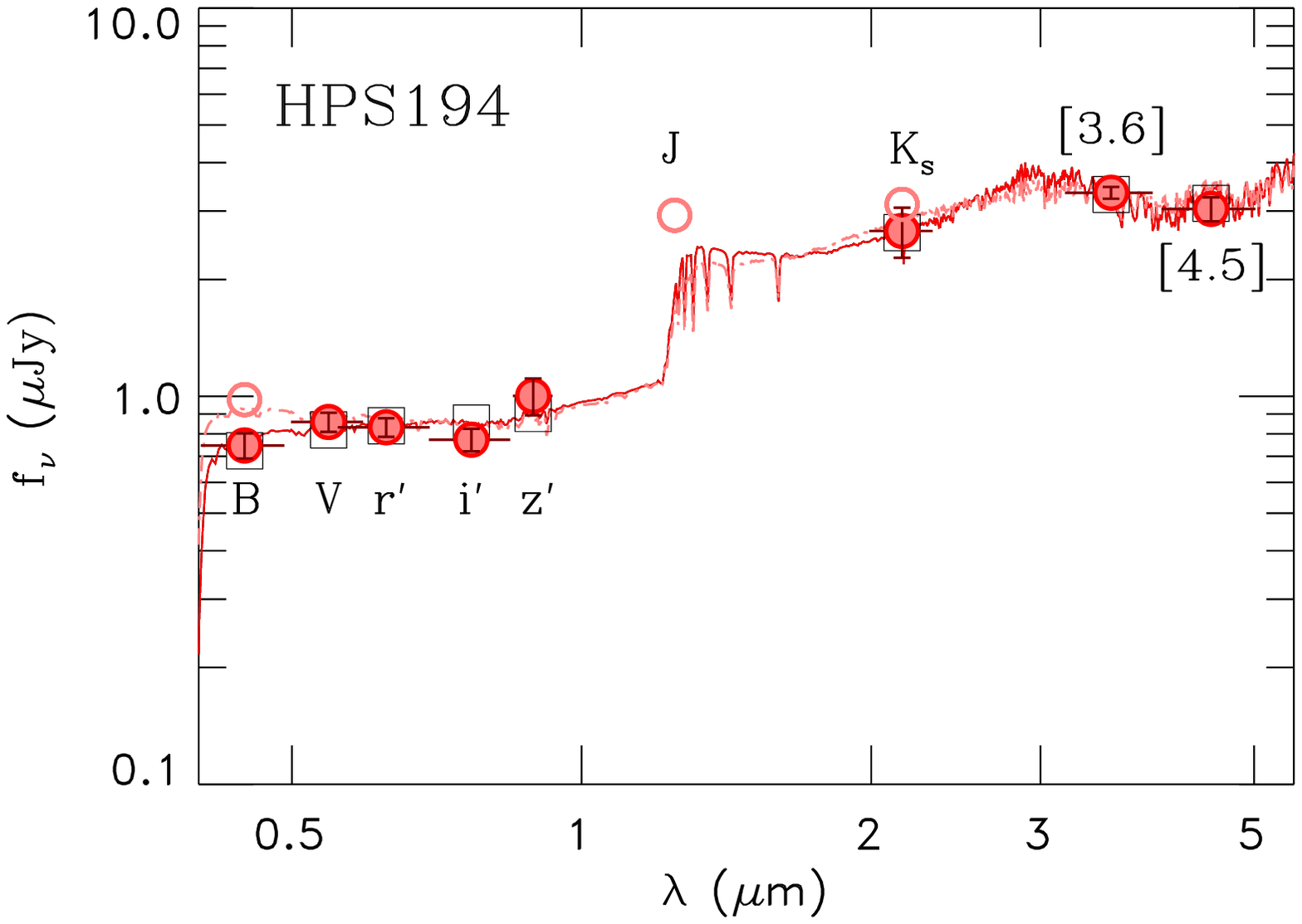}{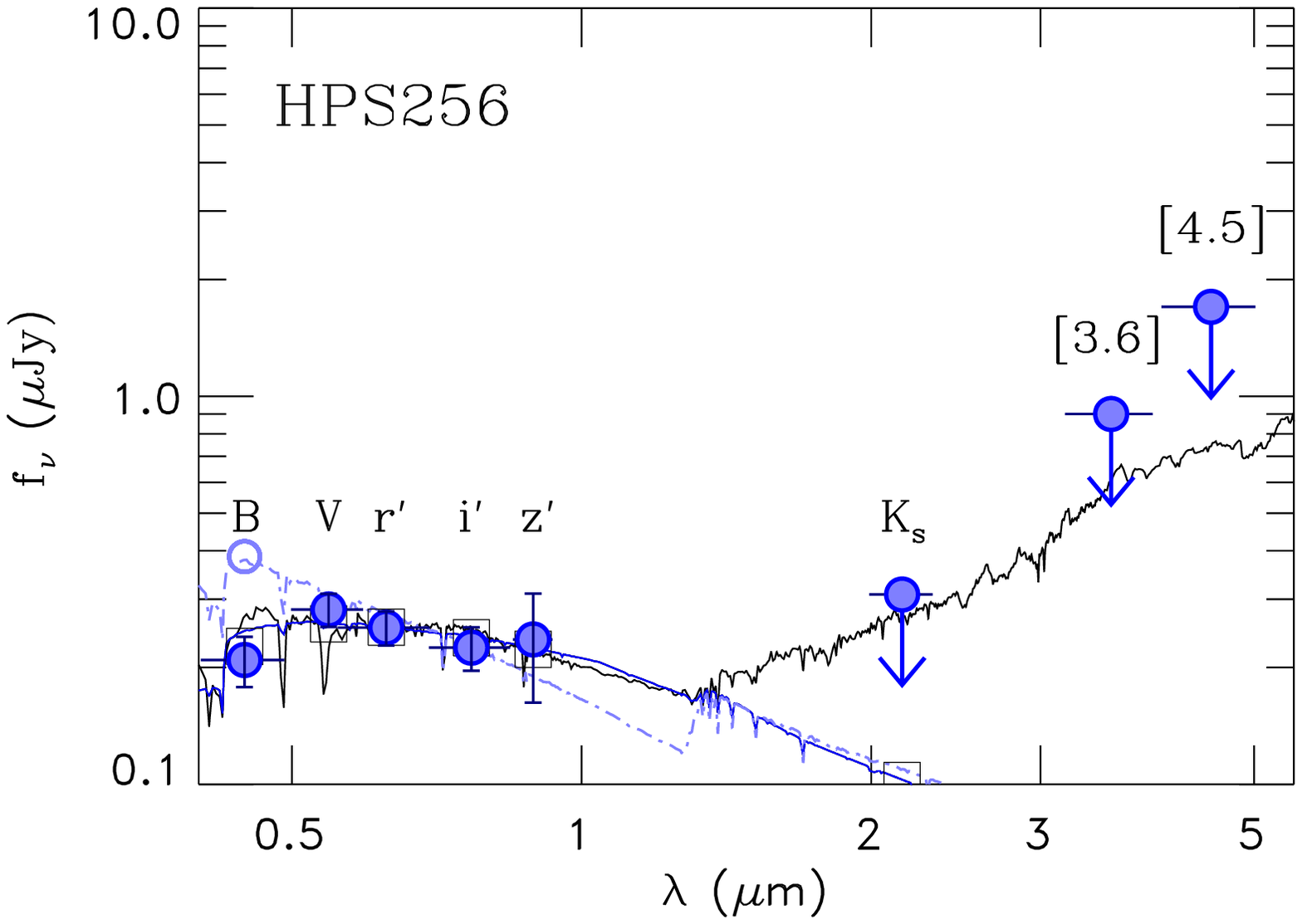}
%\vspace{6mm}
\caption{The results from performing SED-fitting on HPS194 (left; in red) and HPS256 (right; in blue).  The filled circles show the observed fluxes, with the open B and K$_{s}$-band points denoting the fluxes in those bands prior to subtraction of the emission lines (an open circle also denotes the measured J-band flux of HPS194, which we exclude from the emission-line corrected fit, as it may be contaminated by [O\,{\sc ii}] emission).  The light-colored line shows the best-fit when emission line contamination is ignored, and the dark-colored line shows the best-fit model when the measured \lya\ and H$\alpha$ emission line fluxes have been subtracted from the B and K$_{s}$-band fluxes.  The squares show the bandpass-averaged flux of the best-fit emission-line corrected model.  For HPS256, we show the 1$\sigma$ upper limit for the K-band flux, and the 5$\sigma$ upper limits for the IRAC fluxes (as only objects brighter than this were included in the S-COSMOS catalog).   In both objects differences arise when the emission line fluxes are removed.  Specifically, in HPS194 a redder UV slope is allowed after the subtraction of \lya\, which results in lower value of $\tau_{SFH}$ and a higher extinction.  Combining this with the lowered K$_{s}$-band flux results in a much younger best-fit age.  The black curve in the right-hand panel shows the best-fit maximal mass model, which has a mass $\sim$ 100$\times$ greater than the single-population model, as this object is unconstrained at $\lambda > 1 \mu$m.}\label{sedfit}
\end{figure*}

The SED-fitting results are tabulated in Table 4, and the best-fit models are shown in \fig{sedfit}.  These two LAEs appear to be very different physically.  HPS194 appears very old, near the age of the universe at $z =$ 2.29, and is very massive, with a stellar mass of $\sim$ 3 $\times$ 10$^{10}$ M\sol.  HPS256 appears very young, with an age of only a few Myr, and is more than two orders of magnitude less massive.  Both objects have very low stellar metallicity values of $\leq$ 0.02 $Z$\sol, in agreement with the upper limits on their gas-phase metallicities.

However, from our observed optical and NIR spectra, it is apparent that the emission lines we have measured in these objects are very strong, and they may be adversely affecting the model results.  In order to quantify this effect, we performed a second iteration of model fitting, where we corrected the broadband photometry for the observed \lya\ and H$\alpha$ emission-line fluxes.  This was done in the following steps, following the appendix of \citet{guaita10}.  First, we computed the ratio between the filter transmission at $\lambda_{line}$ and the maximum of the transmission function, R$_{T}$.  The amount of flux that the line contributes to the broadband measurement was computed as
\begin{equation}
f_{\nu, line} = R_{T} \times \frac{F_{line}}{\int T(\lambda) \frac{c}{\lambda^{2}}d\lambda},
\end{equation}
\noindent where T($\lambda$) is the filter transmission function.  This flux was subtracted from the catalog photometry.  This correction was made for the $B$-band for \lya\, accounting for a 0.30 mag correction for HPS194, and a 0.67 mag correction for HPS256.  The correction due to H$\alpha$ was made to the K$_{s}$-band for HPS194, and accounted for a 0.18 mag correction.  Additionally, we also excluded the J-band flux from the fit in HPS194 (by making its uncertainty very large), as it appears to contain an excess in \fig{sedfit}, which could be due to strong [O\,{\sc ii}] emission, though future J-band spectroscopy is needed to confirm this suggestion.  The correction due to the \lya\ fluxes will primarily affect the UV slope, and thus the derived dust extinction, while the correction due to H$\alpha$ will change the amplitude of the $z - K_{s}$ color and the derived 4000 \AA\ break, and thus the derived stellar population age.
\begin{deluxetable*}{ccccccccccc}
\tablecaption{Derived Physical Properties from SED Ftting}
\tablewidth{0pt}
\tablehead{
\colhead{Object} & \colhead{Mass} & \colhead{Mass} & \colhead{Age} & \colhead{Age} & \colhead{$A_\mathrm{V}$} & \colhead{$A_\mathrm{V}$} & \colhead{$Z$} & \colhead{$Z$} & \colhead{$\tau_{SFR}$} & \colhead{$\tau_{SFR}$}\\
\colhead{$ $} & \colhead{Best Fit} & \colhead{68\% Range} & \colhead{Best Fit} & \colhead{68\% Range} & \colhead{Best Fit} & \colhead{68\% Range} & \colhead{Best Fit} & \colhead{68\% Range} & \colhead{Best Fit} & \colhead{68\% Range}\\
\colhead{$ $} & \colhead{(10$^{9}$ $M$\sol)} & \colhead{(10$^{9}$ $M$\sol)} & \colhead{(Myr)} & \colhead{(Myr)} & \colhead{(mag)} & \colhead{(mag)} & \colhead{($Z$\sol)} & \colhead{($Z$\sol)} & \colhead{(yr)} & \colhead{(yr)}\\
}
\startdata
HPS194\phantom{ Corrected}&30.1&28.0 -- 31.5&2500&2400 -- 2600&0.08&0.00 -- 0.08&0.02&0.02 -- 0.02&10$^{9.6}$&10$^{9.6}$ -- 10$^{9.6}$\\
\smallskip
HPS194 Corrected&15.1&10.3 -- 28.7&290&140 -- 2200&0.24&0.00 -- 0.16&0.02&0.02 -- 0.20&10$^{8}$&10$^{6}$ -- 10$^{9.6}$\\
HPS256\phantom{ Corrected}&\phantom{0}0.10&0.06 -- 0.10&3&2 -- 3&0.00&0.00 -- 0.00&0.02&0.02 -- 0.40&10$^{5}$&10$^{5}$ -- 10$^{5}$\\
HPS256 Corrected&\phantom{0}0.59&0.28 -- 1.42&2&1 -- 57&0.81&0.00 -- 0.97&0.02&0.02 -- 0.02&10$^{5}$&10$^{5}$ -- 10$^{7}$\\
\enddata
\tablecomments{The best-fit physical properties from our SED-fitting analysis, along with the 68\% confidence range on each property, measured from Monte Carlo simulations.  The first row for each object shows the best-fit values for the catalog photometry, and the second row presents the results when correcting the photometry for the measured \lya\ and H$\alpha$ emission line fluxes.}
\end{deluxetable*}

The results from the SED-fitting iteration with the corrected broadband fluxes are also given in Table 4, and are shown in \fig{sedfit}.  The corrected fit for HPS256 is $\sim$ 6$\times$ more massive, and though their ages are similar, the uncertainty on the age in the emission-line corrected model extends to older ages.  However, as this galaxy is not detected in its rest-frame optical, this fit is relatively unconstrained.  Specifically, a single population fit cannot constrain the amount of mass in old stars.  To examine this uncertainty, we performed a two-population fit, where the second population composes 96\% of the stellar mass, and is formed in a burst at $z=$20.  This model had a best-fit mass of 1.3 $\times$ 10$^{10}$ M\sol, which is the maximum possible mass in this galaxy given the flux limits at longer wavelengths, and is a factor of $>$ 100$\times$ greater than the stellar mass of the single-population fit.  We show this model as the black curve in the right-hand panel of \fig{sedfit}.

The changes to the physical properties of HPS194 are more significant.  Without accounting for the emission lines, this object appeared to have an age of $\sim$ 2.5 Gyr.  Subtracting the measured emission line flux from the observed photometry drastically reduces the best-fit age to $\sim$ 300 Myr.  Additionally, the stellar mass is a factor of $\sim$ 2$\times$ lower, and the derived dust extinction increases.  This change can be understood when examining \fig{sedfit}.  The reduction in the flux of the $B$-band results in a redder fit to the UV slope, implying a greater level of dust extinction\footnote[9]{Although SED fitting shows that HPS194 appears to have A$_\mathrm{V}$ $\sim$ 0.25 mag, this is consistent with the Balmer decrement measurement, which allows a range of 0 $\lesssim$ A$_\mathrm{V}$ $\lesssim$ 1, as well as the UV-slope derived value of A$_\mathrm{V}$ = 0.4 $\pm$ 0.2 from Blanc et al.\ (2010).}.  At the same time, the lessened K$_{s}$-band flux lowers the z$^{\prime}$ $-$ K$_{s}$ color.  This, combined with the increased dust extinction, requires a much lower age to fit the observed z$^{\prime}$ - [3.6] color.  While the difference in the best-fit models is substantial, the difference in mass and age between the two is only at the $\sim$ 2$\sigma$ level, primarily because these objects at $z \sim$ 2.3 are somewhat evolved, containing significant mass in old stars.  At very high redshifts, where galaxies are less evolved, and there are fewer photometric points to constrain the SEDs, emission line contamination likely plays a much larger role.

%The correction to the u$^{\ast}$ and B-band fluxes due to the \lya\ emission does have some inherent uncertainty, as the IGM is attenuating flux in the same bands which are affected by the \lya\ emission.  To examine this effect, we ran a final iteration of our model fitting, using only bands redward of the B-band (and thus redward of \lya).  We found a similar range of allowable properties for HPS256.  For HPS194, the stellar mass was still constrained to be $\sim$ 3$\times$ lower than the uncorrected fit, while the stellar age was found to be $\sim$ 156 Myr, thus somewhat older than when the \lya-corrected bands were included, though still much younger then when the \lya\ emission is not corrected.  However, the 68\% confidence range on the age is larger, from 130 -- 1800 Myr, with slightly overlaps with the 68\% age range from the uncorrected fit.  We conclude that while the uncertainties on the best-fit age can be large, ignoring the presence of emission lines in this object would cause an incorrect estimate of its age, especially if the uncertainties on the age are not considered.

\subsection{Implications for Very High Redshifts}
The equivalent widths of our lines are difficult to measure from the spectra alone, as in both cases the continuum is not detected.  The \lya\ EWs are computed by deriving the continuum flux via interpolation of the flux from all detected optical filters redward of \lya, as discussed in Paper I.  For the rest-frame optical emission lines, we use the continuum flux of the best-fit emission-line corrected model to compute the EWs, where EW = f$_{line}$ / f$_{continuum}$, where f$_{continuum}$ is the continuum flux (in f$_{\lambda}$ units) of the model at the wavelength of the emission line.  While the continuum flux of HPS194 is well-constrained at these wavelengths due to the IRAC detections, the same is not true for HPS256, thus only EWs for HPS194 are shown in Table 2.  The rest-frame EWs of both \OIII\ lines and H$\alpha$ are very large, with EW $\sim$ 200 \AA\ for both \OIII\ $\lambda$5007 and H$\alpha$ in HPS194.

The contributions of emission lines such as these to the integrated fluxes of galaxies will have an even greater effect at high redshift, especially in the rest-frame optical.  The amount that the emission line EWs affect the broadband photometry increases with redshift, as
\begin{equation}
\Delta m \simeq -2.5 \mathrm{log} \left[1 + \frac{W_\mathrm{0}(1 + z)}{\Delta \lambda}\right],
\end{equation}
due to the increase of the observed EW with $1 + z$ (or alternatively, the shrinking of the FWHM of a filter in the rest frame).  At $z \sim$ 7, \citet{labbe10} stacked a sample of 12 z-dropout galaxies, and detect the stack in both IRAC 3.6 and 4.5 $\mu$m bands.  The resulting stellar population fit implies a strong 4000 \AA\ break, due to the $H$ $-$ 3.6 $\mu$m color.  The galaxies used in the stack have photometric redshifts from $\sim$ 6.5 -- 7.5, thus the \OIII\ $\lambda\lambda$4959,5007 emission lines can contaminate the fluxes in both of these IRAC bands.  The total EW of both \OIII\ lines measured in HPS194 is $\sim$ 270 \AA.  Assuming the objects are evenly distributed in redshift, this will place $\sim$ half of the line flux in each IRAC band.  A EW $\sim$ 140 \AA\ emission line at $z \sim$ 7 in both IRAC bands will increase the IRAC fluxes, according to Equation 1, by $\Delta$m$_{3.6}$ = 0.15 and $\Delta$m$_{4.5}$ = 0.11 mag.  This reduction in the amplitude of the 4000 \AA\ break of the best-fit model will result in a younger best-fit age.  This effect is crucial, as \citet{labbe10} find a best-fit age of 300 Myr, which implies that the stars in these galaxies began forming at $z \sim$ 11.  In fact, including theoretical nebular emission lines in their stellar population models, \citet{schaerer10} find a best-fit age to the photometry of \citet{labbe10} of only 4 Myr \citep[see also][]{finlator10, watson10}.

While measurements of objects in the mid-infrared at high redshift are currently very challenging, the {\it James Webb Space Telescope} will soon open up a wide range of high-redshift objects for SED analyses, thus the important role of emission lines needs to be accounted for to ensure accurate stellar population results.

\section{Conclusions}
We report detections of multiple rest-frame optical emission lines from high-redshift \lya-selected galaxies.  We detect H$\alpha$, H$\beta$, and \OIII\ from two galaxies at $z =$ 2.3 and $z =$ 2.5, discovered with the HETDEX pilot survey.  We have used these emission line measurements to spectroscopically probe the physical properties of LAEs for the first time.

Using the redshifts of the rest-frame optical emission lines as a measure of the systemic redshift, we find differences between the redshift of \lya\ emission and the systemic redshift of $\Delta$v = $+$162 $\pm$ 37 (photometric) $\pm$ 42 (systematic) km s$^{-1}$ for HPS194, and $\Delta$v = $+$36 $\pm$ 35 $\pm$ 18 km s$^{-1}$ for HPS256.  The slightly higher redshift of \lya\ is similar to what has been seen in more evolved high-redshift galaxies, and is potentially due to a large-scale outflow in the ISMs of these galaxies.  Combining our results with the only two other LAEs which have had their velocity offsets measured, we find that 3/4 of them exhibit outflows, with most at velocities less than are typically seen in LBGs ($>$ 200 km s$^{-1}$).  This may be due to lower star formation rates in the LAEs, though a larger sample is needed to clearly correlate the velocity offsets in LAEs with any physical property.

Using the measured emission line flux ratios (and limits), we probe for the presence of AGN in these LAEs, using the BPT line-ratio diagnostic diagram.  We find that both LAEs lie far from the AGN sequence, consistent with higher-redshift LAEs, in which AGNs are rare.  We use the measured H$\alpha$ and H$\beta$ lines in our objects to measure the dust extinction.  However, as the detection significance of H$\beta$ in both objects is $<$ 10$\sigma$, the uncertainties on the derived extinctions are high, thus we cannot definitively rule out zero dust in these objects.  However, the measured ratio of \lya/H$\alpha$ in both objects is $\sim$ 7, thus some dust is likely causing deviation from the Case B ratio of 8.7, though the geometry of such an ISM is as yet unconstrained.  The escape fractions of \lya\ photons in these galaxies are high, with f$_{esc}$(\lya) $\sim$ 60\% in both LAEs.

Lacking AGN contamination, we can use emission line ratios to measure the gas-phase metallicities.  As the \NII\ line is undetected in both objects, we can only place upper limits on their metallicities, which we find to be $<$ 0.17 and $<$ 0.28 $Z$\sol\ for HPS194 and HPS256, respectively.  Combining these limits with the stellar masses measured from emission-line corrected SED-fitting results, we find that at least one of these objects lies well below the mass-metallicity sequence for LBGs at similar redshifts, implying that objects selected on the basis of their \lya\ emission, even if they are relatively massive, may be more metal poor than the general galaxy population at $z \sim$ 2.3.

Finally, we examine the results of SED fitting with and without correcting for the presence of \lya\ and H$\alpha$ emission lines.  We find that ignoring this strong line emission can result in an overestimation of the age by orders of magnitude, and an over-estimation of the stellar mass by a factor of two.  This is highly significant for high-redshift studies, as the rest-frame EWs of both \OIII\ and H$\alpha$ are $\sim$ 200 \AA.  The amount that an emission line affects a broadband flux is proportional to $1+z$, thus correcting for this contamination is crucial at high redshift.

Near-infrared spectroscopy allows one to probe the physical characteristics of high-redshift galaxies with a much higher precision than is possible with SED-fitting techniques alone.  Although it is currently time expensive, the presence of multi-object NIR spectrographs will alleviate this in the near future.  The ability of the {\it James Webb Space Telescope} to obtain mid-infrared spectroscopy will allow these analyses to be pushed to higher redshifts.

\acknowledgements

The authors wish to recognize and acknowledge the very significant cultural role and reverence that the summit of Mauna Kea has always had within the indigenous Hawaiian community.  We are most fortunate to have the opportunity to conduct observations from this mountain.  The construction of VIRUS-P was possible thanks to the generous support of the Cynthia \& George Mitchell Foundation.  We thank Darren DePoy for stimulating conversations.  SLF was supported by the Texas A\&M Department of Physics and Astronomy.  We acknowledge funding from the Texas Norman Hackerman Advanced Research Program under grants ARP 003658-0005-2006 and 003658-0295-2007. JJA acknowledges the support of a National Science Foundation Graduate Research Fellowship during this work.

\end{document}